\documentclass[english, 11pt, a4paper]{article}

\usepackage{slashed}
\usepackage[utf8]{inputenc}
\usepackage{babel}
\usepackage{graphics}
\usepackage{graphicx}
\usepackage{texdraw}
\usepackage{epsfig}
\usepackage{float}
\usepackage{amsmath}
\usepackage{amssymb}
\usepackage{enumerate}
\usepackage[all]{xy}
\usepackage{fullpage}
\usepackage[small]{caption}
\usepackage{hyperref}
\usepackage{physics}

\newcommand{\be}{\begin{equation}}
\newcommand{\ee}{\end{equation}}

\begin{document}

\title{{\bf Axial $U_A(1)$ Anomaly: a New Mechanism to Generate Massless Bosons}}

\date{}
\author{Vicente Azcoiti \\
        Departamento de F{\'i}sica Te\'orica, Facultad de Ciencias, and \\
        Centro de Astropartículas y Física de Altas Energías (CAPA),\\ Universidad de Zaragoza,
        Pedro Cerbuna 9, 50009 Zaragoza, Spain\\}

\maketitle

\begin{abstract}

  Prior to the establishment of $QCD$ as the correct theory describing hadronic physics, it was
  realized that the essential ingredients of the hadronic world at low energies are chiral
  symmetry and its spontaneous breaking. Spontaneous symmetry breaking is a non-perturbative
  phenomenon and thanks to massive $QCD$ simulations on the lattice we have at present a good
  understanding on the vacuum realization of the non-abelian chiral symmetry as a function of
  the physical temperature. As far as the $U_A(1)$ anomaly is concerned, and
  especially in the high temperature phase, the current situation is however
  far from satisfactory. 
  The first part of this article is devoted to review the present status of lattice calculations, 
  in the high temperature phase of $QCD$, of quantities directly related 
  to the $U_A(1)$ axial anomaly. In the second part I will analyze some interesting physical
  implications of the $U_A(1)$ anomaly, recently suggested, in systems where the non-abelian
  axial symmetry is
  fulfilled in the vacuum. More precisely I will argue that, if the $U_A(1)$ symmetry 
  remains effectively broken, the topological properties of the theory can be the basis of a
  mechanism, other than Goldstone's theorem, to generate a rich spectrum of massless bosons
  at the chiral limit.
  \vskip 1cm
  \noindent
  {\bf Keywords:} Chiral Transition, Lattice $QCD$, $U(1)$ Anomaly, Topology, Massless Bosons
\end{abstract}

\vfill\eject

\section{Introduction} \label{Introduction}

Nowadays we know that symmetries play an important role in determining the Lagrangian of a
quantum field theory. There are essentially two types of symmetry, local ones or gauge
symmetries, and global ones. The gauge symmetries are characterized by transformations which
depend on the space-time coordinates while in global symmetries the transformations are
space-time independent. Also gauge symmetries serve to fix the couplings of the Lagrangian, and
global symmetries allow us to to assign quantum numbers to the particles and to predict
the existence of massless bosons when a continuous global symmetry is
spontaneously broken.

In what concerns $QCD$, the theory of the strong interaction, and prior to the establishment of
this theory as the correct theory describing hadronic physics, it was
realized that the essential ingredients of the hadronic world at low energies are chiral symmetry
and its spontaneous breaking. Indeed these two properties of the strong interaction have
important phenomenological implications, and allow us to understand some puzzling phenomena
as why pions have much smaller masses than the the proton mass,
and why we do not see degenerate masses for chiral partners in the boson sector, and parity
partners in the baryon sector.

Chiral symmetry breaking by the vacuum state of QCD is a nonperturbative
phenomenon, that results from the interaction of many microscopic degrees of
freedom, and which can be investigated mainly through lattice QCD simulations. As a matter of
fact, lattice $QCD$ is the most powerful technique for investigating non-perturbative effects from
first principles. However, putting chiral symmetry onto the lattice turned out to be
a difficult task. The underlying reason is that a naive lattice regularization suffers from the
doubling problem. The addition of the Wilson term to the naive action solves the doubling
problem but breaks chiral symmetry explicitly, even for massless quarks. This is usually not
considered to be a fundamental problem because we expect that the symmetry is restored in the
continuum limit. However, at finite lattice spacing, chiral symmetry may still be rather
strongly violated by lattice effects.

On the other hand, staggered fermions cope to the doubling problem reducing the number of species
from sixteen to four, and to reduce the number of fermion species from four to one, a
rooting procedure has been used. Even if controversial, the rooting procedure has allowed
to obtain very accurate results in lattice $QCD$ simulations with two and three flavors.

The doubling problem cannot be simply overcome because there is a fundamental theorem by Nielsen
and Ninomiya which states that, on the lattice, one cannot implement
chiral symmetry as in the continuum formulation, and at the same time have a theory free
of doublers. However, despite this difficulty, the problem of chiral symmetry on the lattice
was solved at the end of the past century with a generalization
of chiral symmetry, through the so-called Ginsparg–Wilson equation for the
lattice Dirac operator, which replaces the standard anticommutation relation of the continuum
formulation $D\gamma_5 + \gamma_5 D = 0$ by $D\gamma_5 + \gamma_5 D = a D\gamma_5 D$.
With this new concept a clean implementation of chiral symmetry on the lattice has been
achieved. The axial transformations reduce to the continuum transformations in the naive
continuum limit, but at finite lattice spacing, $a$, an axial transformation involves also the
gauge fields, and this is how Ginsparg-Wilson's formulation evades Nielsen-Ninomiya theorem.

All these features are well established in the lattice community, and the interested reader can
find in \cite{langlib}, for instance, a very good guide.

Returning to the topic of $QCD$ phenomenology, there is also another puzzling phenomenon which is
known as the $U(1)$ problem. The $QCD$ Lagrangian for massless quarks is invariant under the
chiral group $U_V(N_f)\times U_A(N_f) = SU_V(N_f)\times SU_A(N_f) \times U_V(1) \times U_A(1)$,
with $V$ and $A$ denoting vector and axial vector transformations respectively. Below $1 GeV$ 
the flavor index $f$ runs from 1 to 3 (up, down and strange quarks), and the chiral symmetry group
is $U_V(3)\times U_A(3)$. The lightweight pseudoscalars found in Nature suggest, as stated before,
that the $U_A(3)$ axial symmetry is spontaneously broken in the chiral limit, but in such a case
we would have nine Goldstone bosons. The pions, $K$-meson, and $\eta$-meson are eight of them but
the candidate for the ninth Goldstone boson, the $\eta'$-meson, has too great a mass to be a
quasi-Goldstone boson. This is the axial $U(1)$ problem that ’t Hooft solved by realizing that the
$U_A(1)$ axial symmetry is anomalous at the quantum level. ’t Hooft’s resolution of the U(1)
problem suggests in a natural way the introduction of a $CP$ violating term in the $QCD$
Lagrangian, the $\theta$-term, thus generating another long standing problem, the strong $CP$
problem.

Thanks to massive QCD simulations on the lattice, we have at present a good qualitative and
quantitative understanding on the vacuum realization of the non-abelian $SU_A(N_f)$ chiral
symmetry, as a function of the physical temperature, but as far as $U_A(1)$ anomaly and its
associated $\theta$ parameter
are concerned, and especially in the high temperature
phase, the current situation is far from satisfactory, and this makes understanding
the role of the $\theta$ parameter in QCD, and its connection with the strong CP
problem, one of the biggest challenges for high energy theorists \cite{peccei}.

The aim to elucidate the existence of new low-mass weakly
interacting particles from a theoretical, phenomenological, and experimental point of view is
intimately related to this issue.
The light particle that has gathered the most attention has been the axion, predicted 
by Weinberg \cite{weinberg} and Wilczek \cite{wilczek}, in the Peccei and Quinn
mechanism \cite{pq}, to explain the absence
of parity and temporal invariance violations induced by the QCD vacuum. The axion is one of
the more interesting candidates to make the dark matter of the universe, and
the axion potential, that determines the dynamics of the axion field, plays a fundamental
role in this context.

The calculation of the
topological susceptibility in QCD is already a challenge, but calculating the complete
potential requires a strategy to deal with the so called sign problem, that is,
the presence of a highly oscillating term in the path integral.
Indeed, Euclidean lattice gauge theory has not been able to help us much because of the imaginary
contribution to
the action, coming from the $\theta$-term, that prevents the applicability of the importance
sampling method \cite{vicari}.

The $QCD$ axion model relates the topological susceptibility $\chi_T$ at $\theta=0$ with
the axion mass $m_a$ and decay constant $f_a$ through the relation $\chi_T = m^2_a f^2_a$.
The axion mass is, on the other hand, an essential ingredient in the calculation of the
axion abundance in the Universe. Therefore a precise computation of the temperature
dependence of the topological susceptibility in $QCD$ becomes of primordial interest in
this context.

In this article I will focus on the current status of the lattice calculations, 
in the high temperature chirally symmetric phase of $QCD$, of quantities directly related 
to the $U_A(1)$ axial anomaly, as the topological and axial $U_A(1)$ susceptibilities, and
screening masses, and will 
also discuss on some interesting physical implications of the $U_A(1)$ axial anomaly in systems
where the non-abelian axial symmetry is fulfilled in the vacuum. I will briefly review in
section \ref{review} some theoretical prejudices about the effects of the axial anomaly in the
high temperature phase of $QCD$, and will analyze what the results of the numerical simulations
on the lattice suggest on the effectiveness of the axial anomaly in this phase. In  
section \ref{The main} I will argue that the topological properties of a quantum field theory, 
with $U_A(1)$ anomaly and exact non-abelian axial symmetry, 
as for instance $QCD$ in the high temperature phase, can be the basis of a mechanism,
other than Goldstone's theorem, to generate a rich spectrum of massless bosons at the
chiral limit. 
The two-flavor Schwinger model, which was analyzed by Coleman \cite{coleman} many years ago,
is an excellent test bed for verifying the predictions of section \ref{The main}, and
section \ref{Schwinger} contains the results of this test. The last section contains a 
discussion of the results reported in this article.

\section{Theoretical biases versus numerical results} \label{review}

The large mass of the $\eta'$ meson should come from the effects of the $U_A(1)$ axial anomaly
and its related gauge field topology, both present in $QCD$. Despite the difficulty of computing
the contribution of disconnected diagrams to the $\eta'$ correlator in lattice simulations,
these obstacles have been overcome and lattice 
calculations \cite{etauno}, \cite{etados}, \cite{etatres} give a mass for
the $\eta'$ meson compatible with its experimental value, and this can be seen as an indirect
confirmation that the effects of the anomaly are present in the low temperature phase of QCD. 

Conversely, the current situation regarding the fate of the axial anomaly in the high temperature
phase of $QCD$, where the non-abelian axial symmetry is not spontaneously broken, is unclear,
and this is quite unsatisfactory. The nature of the chiral phase transition in two-flavor $QCD$,
for instance, is affected by the way in which the effects of the $U_A(1)$ axial anomaly
manifest themselves around the critical temperature \cite{wilczek2}. Indeed, if the $U_A(1)$
axial symmetry remains effectively broken, we expect a continuous chiral transition belonging to 
the three-dimensional $O(4)$ vector universality class, which shows a critical
exponent $\delta = 4.789(6)$ \cite{hasenbu}, while if $U_A(1)$ is effectively restored,
the chiral transition is first order, or second order with critical exponents belonging to the
$U_V(2)\times U_A(2)\rightarrow U_V(2)$ universality class ($\delta = 4.3(1)$) \cite{peliss}.

The first investigations on the fate of the $U_A(1)$ axial anomaly in the chiral symmetry
restored phase of $QCD$ started a long time ago. The idea that the chiral symmetry restored phase
of two-flavor QCD could be symmetric under $U_V(2)\times U_A(2)$ 
rather than $SU_V(2)\times SU_A(2)$ was raised by Shuryak in 1994 \cite{shu}, based on an
instanton liquid-model study. In 1996 Cohen \cite{cohen1} showed, using the
continuum formulation of two flavor QCD, and assuming the absence of the zero mode’s
contribution, that all the disconnected contributions to the two-point correlation functions in
the $SU_A(2)$ symmetric phase at high temperature vanish in the chiral limit. The main conclusion
of this work was that the eight scalar and pseudoscalar mesons should have the same mass in the
chiral limit, the typical effects of the $U_A(1)$ axial anomaly being absent in this
phase. Also Cohen argued in \cite{cohen2} that the analyticity of the free energy density in the
quark mass $m$, around $m=0$, in the high temperature phase, imposes constraints on the
spectral density of the Dirac operator around the origin which are enough to guarantee the
previous results.

Later on, Aoki et al. \cite{diez} got constraints on the Dirac spectrum of
overlap fermions, strong enough for all of the
$U(1)_A$ breaking effects among correlation functions of scalar and pseudoscalar
operators to vanish, and they concluded that there is no remnant of the $U(1)_A$ anomaly above
the critical temperature in two flavor $QCD$, at least in these correlation functions. Their
results were obtained under the assumptions that $m$-independent observables are analytic
functions of the square quark-mass $m^2$, at $m=0$, and that the Dirac spectral density can
be expanded in Taylor series near the origin, with a non-vanishing radius of convergence.

The range of applicability of the assumptions made in \cite{diez} is however unclear. 
As stated by the authors, their result strongly relies on their assumption that the vacuum
expectation values of quark-mass independent observables, as the topological susceptibility, are
analytic functions of the square quark-mass, $m^2$, if the non-abelian chiral symmetry is
restored. The two-flavor Schwinger model has a non spontaneously broken $SU_A(2)$ chiral
symmetry and $U_A(1)$ axial anomaly, and Coleman's result for the topological susceptibility in
this model \cite{coleman}

$$
\chi_T\propto m^\frac{4}{3}e^\frac{2}{3}
$$
shows explicitly a non-analytic quark-mass dependence, and thus casts doubt on the general validity
of the assumptions made in \cite{diez}.

In Ref. \cite{trece} a Ginsparg-Wilson fermion lattice regularization was used, and it was
argued that if the vacuum energy density is an analytical
function of the quark mass in the high temperature phase of two-flavor $QCD$,  all
effects of the axial anomaly should disappear. The main conclusion of \cite{trece}
was that either the typical effects 
of the axial $U_A(1)$ anomaly disappear in the symmetric high temperature phase, or the
vacuum energy density shows a singular behavior in the quark mass at the chiral limit.

On the other hand, an analysis of chiral and $U_A(1)$ symmetry restoration based on Ward
identities and $U(3)$ chiral perturbation theory has been carried out in
\cite{jacobo}, \cite{jacobo2}. 
The authors show in their work that in the limit of exact $O(4)$ restoration, understood in terms
of $\delta - \eta$ partner degeneration, the Ward identities analyzed yield also
$O(4)\times U_A(1)$ restoration in terms of $\pi - \eta$ degeneration, and
the pseudo-critical temperatures for restoration of $O(4)$ and $O(4)\times U_A(1)$ tend 
to coincide in the chiral limit.

The first lattice simulations to investigate the fate of the $U_A(1)$ axial anomaly
\cite{cuatro}, \cite{chandra} also started in the 90s. In Ref. \cite{cuatro} the authors
report results of a numerical simulation
of the two-flavor model with staggered quarks. They compute two order parameters,
$\chi_\pi - \chi_\sigma$ for the $SU_A(2)$ chiral symmetry, and $\chi_\pi - \chi_\delta$ for the
$U_A(1)$ axial symmetry, where $\chi_\pi$, $\chi_\sigma$ and $\chi_\delta$ are the pion, $\sigma$
and $\delta$-meson susceptibilities, and they show evidence for a restoration
of the $SU_V(2)\times SU_A(2)$ chiral symmetry, just above the crossover, but not of the axial
$U_A(1)$ symmetry. Ref. \cite{chandra} contains the results of a similar
calculation in two-flavor $QCD$ using also a staggered fermion lattice regularization. As stated
by the authors, the relatively coarse lattice spacing in their simulations,
$a\sim \frac{1}{3}$ Fermi, does not allow for conclusive results on the effectiveness of the
$U(1)_A$ anomaly.

After these pioneering works, this issue has been extensively investigated
using numerical simulations on the lattice, and Refs. \cite{seis}-\cite{nuevo5} are
representative of that. We will focus from now on the most recently obtained results. 

In Ref. \cite{nuevo2} $(2+1)$-flavor $QCD$ is simulated, using chiral domain wall fermions,
for temperatures between $139$ and $196$ MeV. The light-quark mass is chosen
so that the pion mass is held fixed at a heavier-than-physical $200$ MeV value, while the strange
quark mass is set to its physical value. The authors report results for the chiral condensates,
connected and disconnected susceptibilities and the Dirac eigenvalue spectrum, and find a
pseudocritical temperature $T_c\sim 165$ MeV and clear evidence for $U_A(1)$ symmetry breaking
above $T_c$.

Ref. \cite{sharma} contains also a study of $QCD$ with $(2+1)$-flavors of highly
improved staggered
quarks. The authors investigate  the temperature dependence of the anomalous $U_A(1)$ symmetry
breaking in the high temperature phase, and to this end they employ the overlap Dirac operator, 
exploiting its property of preserving the index theorem even at non-vanishing lattice spacing.
The pion mass is fixed to $160$ MeV, and by quantifying the contribution of the near-zero
eigenmodes to $\chi_\pi-\chi_\delta$, the authors conclude that the anomalous breaking of the
axial symmetry in $QCD$ is still visible in the range $T_c\leqslant T \leqslant 1.5 T_c$.

The thermal transition of $QCD$ with two degenerate light flavors is analyzed in \cite{docebis} 
by lattice simulations, using $O(a)$-improved Wilson quarks and the unimproved Wilson plaquette
action. In this work the authors investigate the strength of the anomalous breaking of
the $U_A(1)$ symmetry in the chiral limit by computing the symmetry restoration pattern of
screening masses in various isovector channels, and to quantify the strength of the
$U_A(1)$-anomaly, they use the difference between scalar and pseudoscalar screening masses.
They conclude that their results  suggest that the $U_A(1)$-breaking is strongly reduced at the
transition temperature, and that this disfavors a chiral transition in the O(4)
universality class.

Results for mesonic screening masses in the temperature range $140$ MeV
$\leqslant T \leqslant 2500$ MeV in $(2+1)$-flavor $QCD$, using the highly improved staggered
quark action, are also reported by the $HotQCD$ Collaboration 
in \cite{3f}, with a physical value for the strange quark mass,
and two values of the light quark mass corresponding to pion masses of $160$ and $140$ MeV.
Comparing screening masses for chiral partners, related through the chiral
$SU_L(2)\times SU_R(2)$ and the axial $U_A(1)$ transformations, respectively, the authors 
find, in the case of light-light mesons, evidence for the degeneracy of screening masses related
through the chiral $SU_L(2)\times SU_R(2)$ at or very close to the pseudocritical temperature, 
$T_{pc}$, while screening masses related through an axial $U_A(1)$ transformation start becoming
degenerate only at about $1.3 T_{pc}$.

A recent calculation in $(2 + 1)$-flavor $QCD$ \cite{ding}, using also the highly improved
staggered quark action, shows, after continuum and chiral extrapolations, that the axial
anomaly remains manifested in 2-point correlation functions of scalar and pseudo-scalar mesons
in the chiral limit, at a temperature of about 1.6 times the chiral phase transition
temperature. The analysis is based on novel relations between the nth-order light quark mass
derivatives of the Dirac eigenvalue spectrum, $\rho(\lambda, m_l)$, and the $(n + 1)$-point
correlations among the eigenvalues of the massless Dirac operator, and
the calculations were carried out at the physical value of the strange quark
mass, three lattice spacings, and light quark masses corresponding to pion masses in the range
$55-160$ MeV.

Ref. \cite{nuevo5} contains the latest results of the JLQCD collaboration. In this work
the authors investigate the fate of the $U_A(1)$ axial anomaly in two-flavor $QCD$ at temperatures
$190–330$ MeV using domain wall fermions, reweighted to overlap fermions, at a lattice spacing 
of $0.07$ fm. They measure the axial 
$U_A(1)$ susceptibility, $\chi_\pi -\chi_\delta$, and examine the degeneracy of $U_A(1)$ partners
in meson and baryon correlators. Their conclusion is that all the data above the critical
temperature indicate that the axial $U_A(1)$ violation is consistent with zero within statistical
errors.

All the results discussed so far mainly refer to the temperature dependence of the axial
susceptibility $ U_A (1) $, screening masses, and related quantities. The topological
susceptibility, $\chi_T$, is another observable that can be useful in investigating the fate
of the axial anomaly in the high-temperature phase of $QCD$, and its dependence on temperature
has also been extensively investigated, \cite{martinelli}, \cite{petre}, \cite{javier},
\cite{guido}, \cite{nuevo5}. 

The authors of Ref. \cite{martinelli} explore $N_f = 2+1$ $QCD$ in a  range of temperatures,
from $T_c$ to around $4T_c$, and their results for the topological susceptibility differ strongly,
both in the size and in the temperature dependence, from the dilute instanton gas prediction,
giving rise to a shift of the axion dark-matter window of almost one order of magnitude with
respect to the instanton computation.

The authors of Ref. \cite{petre}, however, observe in the same model very distinct temperature
dependences of the topological susceptibility in the ranges above and below
$250$ MeV; though for temperatures above $250$ MeV, the dependence is found to be consistent
with the dilute instanton gas approximation, at lower temperatures the falloff of topological
susceptibility is milder.

On the other hand, a novel approach is proposed in Ref. \cite{javier}, i.e., the
fixed $Q$ integration, based on the computation of the mean value of the gauge action and chiral
condensate at fixed topological charge $Q$; they find a topological susceptibility
many orders of magnitude smaller than that of Ref. \cite{martinelli} in the cosmologically
relevant temperature region.

A more recent lattice calculation \cite{guido} of the topological properties of $N_f = 2+1$ QCD
with physical quark masses and temperatures around $500$ MeV gives as a result a small but
non-vanishing topological susceptibility, although with large error bars in the continuum limit
extrapolations, pointing that the effects of the $U_A(1)$ axial anomaly still persist at these
temperatures. 

The JLQCD collaboration \cite{nuevo5} reports also results for the topological susceptibility in
two-flavor $QCD$, in the temperature range $195-330$ MeV, for several quark masses, and their data
show a suppression of $\chi_T(m)$ near the chiral limit. The authors claim that their results are
not accurate enough to determine whether $\chi_T (m)$ vanishes at a finite quark mass.

In short we see how, despite the great effort devoted to investigating the fate of the axial
anomaly in the chirally symmetric phase of $QCD$, the current situation on this issue is far from
satisfactory.

\section{Physical effects of the $U_A(1)$ anomaly in models with
  exact $SU_A(N_f)$ chiral symmetry} \label{The main}

We will devote the rest of this article mainly to analyze the physical effects of the $U_A(1)$
anomaly in a fermion-gauge theory with two or more flavors, which exhibits an exact $SU_A(N_f)$
chiral symmetry in the chiral limit. However we will also give a quick look to the one flavor
model, and to the multi-flavor model with spontaneous non-abelian chiral symmetry breaking. 
Although many of the results presented here can be found in
Refs. \cite{trece}, \cite{trecebis} and \cite{catorce}, we will make the rest of this article
self-contained for ease of reading.

We will show in this section that a gauge-fermion quantum field theory, with $U_A(1)$ axial
anomaly, and in which the scalar condensate vanishes in the chiral limit because of an exact
non-Abelian $SU_A(2)$ chiral symmetry, should exhibit a singular quark-mass dependence of the
vacuum energy density and a divergent
correlation length in the correlation function of the scalar condensate,  
if the $U_A(1)$ symmetry is effectively broken. On the contrary, if we assume that all
correlation lengths are finite, and hence the vacuum
energy density is an analytical function of the quark mass, we will see that the
vacuum energy density becomes, at least up to second order in the quark masses,
$\theta$-independent. In the 
former case, the non-anomalous Ward-Takahashi (W-T) identities will tell us
that several pseudoscalar correlation functions, those of the
$SU_A(2)$ chiral partners of the flavor singlet scalar meson, should exhibit a divergent
correlation length too. We will also argue that this result can be generalized for any number
of flavors $N_f> 2$.

\subsection{Some background} \label{background}

To begin, let us write the continuum Euclidean action for a vector-like gauge theory with global 
$U_A(1)$ anomaly in the presence of a $\theta$-vacuum term

\begin{equation}
  S = \int \mathrm{d}^dx \left\{\sum^{N_f}_f\bar\psi_f\left(x\right)
  \left(\gamma_\mu D_\mu\left(x\right)+ m_f\right) \psi_f\left(x\right)
  + \frac{1}{4} F^a_{\mu\nu}\left(x\right)F^a_{\mu\nu}\left(x\right)
  + i\theta Q\left(x\right)\right\}
  \label{eulagran}
\end{equation}
where $d$ is the space-time dimensionality, $D_\mu(x)$ the covariant derivative, $N_f$ the
number of flavors, and $Q(x)$ the density of topological charge of the gauge configuration.
The topological charge $Q$ is the integral of the density of topological charge $Q(x)$ over the
space-time volume, and it is an integer number which in the case of $QCD$ reads as follows 

\begin{equation}
  Q = \frac{g^2}{64\pi^2} \int d^4x\epsilon_{\mu\nu\rho\sigma}
  F^a_{\mu\nu}\left(x\right)F^a_{\rho\sigma}\left(x\right).
  \label{ftopcharg}
\end{equation}

To keep mathematical rigor we will avoid ultraviolet divergences with the help of a lattice
regularization, and will use Ginsparg-Wilson (G-W) fermions \cite{Ginsparg},
the overlap fermions \cite{Neuberger1}, \cite{Neuberger2} being an explicit realization of them.
The motivation to use G-W fermions is that they share
with the continuum formulation all essential ingredients. Indeed G-W fermions
show an explicit $U_A(1)$ anomalous symmetry \cite{Luscher}, good chiral properties, a
quantized topological charge, and allow us to establish and exact index
theorem on the lattice \cite{Victor}.

The lattice fermionic action for a massless G-W fermion can be written in a compact form as

\begin{equation}
  S_F = a^d \bar\psi D\psi = a^d \sum_{v, w} \bar\psi\left(v\right) D\left(v, w\right)
  \psi\left(w\right)
  \label{fa}
\end{equation}
where $v$ and $w$ contain site, Dirac and color indices, and $D$, the Dirac-Ginsparg-Wilson
operator, obeys the essential anticommutation equation

\begin{equation}
  D\gamma_5 + \gamma_5D = a D\gamma_5D
  \label{antic}
\end{equation}
$a$ being the lattice spacing.

Action (\ref{fa}) is invariant under the following lattice $U_A(1)$
chiral rotation

\begin{equation}
  \psi\rightarrow e^{i\alpha\gamma_5\left(I-\frac{1}{2}aD\right)}\psi,
  \hskip 1cm \bar\psi\rightarrow
\bar\psi e^{i\alpha\left(I-\frac{1}{2}aD\right)\gamma_5}
  \label{chirot}
\end{equation}
which for $a\rightarrow 0$ reduces to the standard continuum chiral transformation. 
However the integration measure of Grassmann variables is not invariant, and the change of
variables (\ref{chirot}) induces a Jacobian

\begin{equation}
e^{-i 2\alpha \frac{a}{2} tr\left(\gamma_5 D\right)}
  \label{jacobian}
\end{equation}
where

\begin{equation}
\frac{a}{2} tr\left(\gamma_5 D\right) = n_- - n_+ = Q
  \label{topcharge}
\end{equation}
is an integer number, the difference between left-handed and right-handed zero modes, which can
be identified with the topological charge $Q$ of the gauge configuration. Equations
(\ref{jacobian}) and (\ref{topcharge}) show us how Ginsparg-Wilson fermions reproduce the
$U_A(1)$ axial anomaly. 

We can also add a symmetry breaking mass term, $m\bar\psi \left(1-\frac{a}{2}D\right)\psi$ to
action (\ref{fa}), so G–W fermions with mass are described by the fermion action 

\begin{equation}
  S_F = a^d \bar\psi D\psi + a^d m\bar\psi \left(1-\frac{a}{2}D\right)\psi
  \label{fam}
\end{equation}
and it can also be shown that the scalar and pseudoscalar condensates

\begin{equation}
S = \bar\psi \left(1-\frac{a}{2}D\right)\psi \hskip 1cm 
P = i\bar\psi \gamma_5\left(1-\frac{a}{2}D\right)\psi
\label{sapc}
\end{equation}
transform, under the chiral $U_A(1)$ rotations (\ref{chirot}), as a vector, just in the same
way as $\bar\psi\psi$ and $i\bar\psi\gamma_5\psi$ do in the continuum formulation.

In what follows we will use dimensionless fermion fields and a dimensionless
Dirac-Ginsparg-Wilson operator. In such a case
the fermion action for the $N_f$-flavor model is

\begin{equation}
  S_F = \sum^{N_f}_f\left\{\bar\psi_f D\psi_f + m_f\bar\psi
  \left(1-\frac{1}{2}D\right)\psi_f\right\}
  \label{famnf}
\end{equation}
where $m_f$ is the mass of flavor $f$ in lattice units. The partition function of this model, 
in the presence of a $\theta$-vacuum term, can be
written as the sum over all topological sectors, $Q$, of the partition function in each
topological sector times a 
$\theta$-phase factor,

\begin{equation}
Z = \sum_{Q} Z_Q e^{i\theta Q}
\label{zeta}
\end{equation}
where $Q$, which takes integer values, is bounded at finite volume by the number of
degrees of freedom. At large lattice volume the partition function should behave as

\begin{equation}
Z\left(\beta,m_f,\theta\right) = e^{-V E\left(\beta,m_f,\theta\right)}
\label{zetalarge}
\end{equation}
where $E\left(\beta,m_f,\theta\right)$ is the vacuum energy density, $\beta$ the inverse 
gauge coupling, $m_f$ the $f$-flavor mass, 
and $V=V_s\times L_t$ the lattice volume in units of the lattice spacing.

\subsection{$Q=0$ topological sector. The one-flavor model and the multi-flavor model with
spontaneous chiral symmetry breaking} \label{The relevance}

In our analysis of the physical phenomena induced by the topological properties of the theory,
the $Q=0$ topological sector will play an essential role, and because of that we devote
this subsection to review some results concerning the relation between vacuum expectation values
of local and non-local operators computed in the $Q=0$ sector, with their corresponding
values in the full theory, which takes into account the contribution of all topological
sectors. In particular we will show that the vacuum energy density, and the vacuum expectation
value of any finite operator, as for instance local or intensive
operators, computed in the $Q=0$ topological sector, is equal, in the infinite volume limit,
to its corresponding value in the full theory. We will also show that this property is in
general not true for non-local operators, the flavor-singlet pseudoscalar susceptibility being
a paradigmatic example of this. However there are non-local operators, as for instance the second 
order fermion-mass derivatives of the vacuum energy density, the value of which in
the $Q=0$ sector match their corresponding values in the full theory, in the infinite lattice
volume limit.

We will also analyze in this subsection the one-flavor case, as well as the multi-flavor case with
spontaneous chiral symmetry breaking, and will show how, although the
aforementioned properties will imply that the $U_A(1)$ symmetry is spontaneously broken in
the $Q=0$ topological sector, the Goldstone theorem is not realized because the divergence of
the flavor-singlet pseudoscalar susceptibility, in this sector, does not originate from a
divergent correlation length \cite{trece}.

The partition function, and the mean value of any operator $O$, as for instance
the scalar and pseudoscalar condensates, or any correlation function, in the $Q=0$ topological
sector, can be computed respectively as 

\begin{equation}
Z_{Q=0} = \frac{1}{2\pi}\int \mathrm{d}\theta Z(\beta,m_f,\theta)
\label{zq0}
\end{equation}

\begin{equation}
\left< O\right>^{Q=0} = \frac{\int \mathrm{d}\theta \left< O\right>_\theta Z(\beta,m_f,\theta)}
{\int \mathrm{d}\theta Z(\beta,m_f,\theta)}
\label{mascurioso}
\end{equation}
where $\left< O\right>_\theta$, which is the mean value of $O$ computed with the lattice
regularized integration 
measure (\ref{eulagran}), is a function of the inverse gauge coupling $\beta$, flavor masses 
$m_f$, and $\theta$, and we will restrict ourselves to the case in which 
it takes a finite value in the infinite lattice volume limit. 
Since the vacuum energy density (\ref{zetalarge}), as a 
function of $\theta$, has its absolute minimum at $\theta=0$ for non-vanishing fermion 
masses, the following relations hold in the infinite volume limit

\begin{equation}
E_{Q=0}\left(\beta,m_f\right) = E\left(\beta,m_f,\theta\right)_{\theta=0}
\label{eq0}
\end{equation}

\begin{equation}
\left< O\right>^{Q=0} = \left< O\right>_{\theta=0}
\label{mascuriosob}
\end{equation}
where $E_{Q=0}\left(\beta,m_f\right)$ is the vacuum energy density of the $Q=0$ topological 
sector.

Taking in mind these results, let us start with the analysis of the one-flavor model at zero
temperature. The results that follow apply, for instance, to one-flavor $QCD$ in four dimensions or
to the one-flavor Schwinger model.

In the one flavor model the only axial symmetry is an
anomalous $U_A(1)$ symmetry. The standard wisdom on the vacuum structure of
this model in the chiral limit is that it is unique at each given value
of $\theta$, the $\theta$-vacuum. Indeed, the only plausible reason to
have a degenerate vacuum in the chiral limit would be the spontaneous
breakdown of chiral symmetry, but since it is anomalous, actually there is
no symmetry. Furthermore, due to the chiral anomaly, the model shows a mass gap in the chiral
limit, and therefore all correlation lengths are finite in physical units. Since the model is
free from infrared divergences, 
the vacuum energy density can be expanded in powers of the
fermion mass $m_u$, treating the quark mass term as a perturbation \cite{Smilga}.
This expansion will be then an ordinary Taylor series

\begin{equation}
  E \left(\beta,m_u,\theta\right) = E_0\left(\beta\right)
  - \Sigma\left(\beta\right) m_u \cos\theta +O(m_u^2),
  \label{LS}
\end{equation}
giving rise to the following expansions for the scalar and pseudoscalar
condensates

\begin{equation}
  \left<S_u\right> =  -\Sigma\left(\beta\right) \cos\theta +O(m_u)
\end{equation}

\begin{equation}
  \left<P_u\right> =  -\Sigma\left(\beta\right) \sin\theta +O(m_u)
  \label{LSc}
\end{equation}
where $S_u$ and $P_u$ are the scalar and pseudoscalar condensates (\ref{sapc}) normalized by the
lattice volume

\begin{equation}
S_u = \frac{1}{V} \bar\psi \left(1-\frac{1}{2}D\right)\psi \hskip 1cm 
P_u = \frac{i}{V}\bar\psi \gamma_5\left(1-\frac{1}{2}D\right)\psi
\label{sapcud}
\end{equation}

The topological
susceptibility $\chi_T$ is given, on the other hand, by the following expansion

\begin{equation}
  \chi_T = \Sigma\left(\beta\right) m_u \cos\theta +O(m_u^2)
\label{topsus}
\end{equation}

The resolution of the $U_A(1)$ problem is obvious if we set down the W-T identity 
which relates the pseudoscalar susceptibility 
$\chi_\eta = \sum_x \left<P_u\left(x\right)P_u\left(0\right)\right>$, the scalar condensate
$\left<S_u\right>$, and the topological susceptibility $\chi_T$

\begin{equation}
  \chi_\eta =  -\frac{\left<S_u\right>}{m_u} -
  \frac{\chi_T}{m_u^2}.
  \label{Trans-II}
\end{equation}
Indeed the divergence in the chiral limit of the first term in the
right-hand side of (\ref{Trans-II}) is canceled by the divergence of the second term in
this equation, giving rise to a finite pseudoscalar susceptibility, and a finite non-vanishing
mass for the pseudoscalar $\eta$ boson.

In what concerns the $Q=0$ topological sector, we want to notice two relevant features:

\begin{enumerate}
\item
  The global $U_A(1)$ axial symmetry is not anomalous in the $Q=0$ topological sector.
  
\item
  If we apply equation (\ref{mascuriosob}) to the computation of the vacuum expectation value
  of the scalar condensate, we get that the $U_A(1)$ symmetry is spontaneously broken in the
  $Q=0$ sector because the chiral limit of the infinite volume limit of the scalar condensate,
  the limits taken in this order, does not vanish.
 
  \end{enumerate}

Equation (\ref{mascurioso}) allow us to write for the infinite volume limit of the two-point
pseudoscalar correlation function, $\left<P_u\left(x\right)P_u\left(0\right)\right>$, 
the following relation

\begin{equation}
  \left<P_u\left(x\right)P_u\left(0\right)\right>^{Q=0} =
  \left<P_u\left(x\right)P_u\left(0\right)\right>_{\theta=0}.
  \label{mascurioso2}
\end{equation}
This equation implies that the mass of the pseudoscalar boson, $m_\eta$, which can be extracted
from the long distance behavior of the two-point correlation function, computed in the $Q=0$
sector, is equal to the value we should get in the full theory, taking into account the
contribution of all topological sectors. On the other hand the topological susceptibility, 
$\chi_T$, vanishes in the $Q=0$ sector, and hence the W-T identity (\ref{Trans-II} )
in this sector reads as follows

\begin{equation}
  \chi_\eta^{Q=0} = -\frac{\left<S_u\right>^{Q=0}}{m_u}. 
  \label{wardq0}
\end{equation}
This identity tell us that, due to the spontaneous breaking of the $U_A(1)$ symmetry in the
$Q=0$ sector, the pseudoscalar susceptibility 
diverges in the chiral limit, $m_u\rightarrow 0$, in this topological sector. This is a very
surprising result because it suggests that the pseudoscalar boson would be a Goldstone boson,
and therefore its mass, $m_\eta$, would vanish in the $m_u\rightarrow 0$ limit.

The loophole to this paradoxical result is that the divergence of the susceptibility does not
necessarily implies a divergent correlation length. The susceptibility is the infinite volume
limit of the integral of the correlation function over all distances, in this order, and the
infinite volume limit and the space-time integral of the correlation function do not necessarily
commute \cite{trece}. The infinite range interaction Ising model is a paradigmatic example
of the non-commutativity of the two limits.

Let us see with some detail what actually happens. 
The $\left<P_u\left(x\right)P_u\left(0\right)\right>^{Q=0}$ correlation function at any finite
space-time volume $V$ verifies the following equation

\begin{equation}
  \left<P_u\left(x\right)P_u\left(0\right)\right>^{Q=0} = \frac{\int\mathrm{d}\theta 
    \left<P_u\left(x\right)P_u\left(0\right)\right>_{c,\theta} e^{-V E\left(\beta,m,\theta\right)}}
  {\int\mathrm{d}\theta e^{-V E\left(\beta,m,\theta\right)}} + 
  \frac{\int\mathrm{d}\theta
    \left<P_u\left(0\right)\right>^2_{\theta} e^{-V E\left(\beta,m,\theta\right)}}
    {\int\mathrm{d}\theta e^{-V E\left(\beta,m,\theta\right)}}
  \label{menoscurioso}
\end{equation}
where $\left<P_u\left(x\right)P_u\left(0\right)\right>_{c,\theta}$ is the connected pseudoscalar
correlation function at a given $\theta$. The first term in the right-hand side of equation
(\ref{menoscurioso}) converges in the infinite lattice volume limit to
$\left<P_u\left(x\right)P_u\left(0\right)\right>_{\theta=0}$, the pseudoscalar
correlation function at $\theta=0$. In order to compute the large lattice volume behavior of
the second term in the right-hand side of (\ref{menoscurioso}) we can expand
$\left<P_u\left(0\right)\right>^2_{\theta}$, and the vacuum energy density in powers of
the $\theta$ angle as follows

\begin{equation}
  \left<P_u\left(0\right)\right>^2_{\theta} = \left(m_u \chi_\eta +
  \left<S_u\right>\right)^2 \theta^2
  + O(\theta^4).
  \label{expansion}
\end{equation}

\begin{equation}
  E \left(\beta,m_u,\theta\right) = E_{0}\left(\beta,m_u\right)
  - \frac{1} {2} \chi_T\left(\beta,m_u\right)\theta^2 +O(\theta^4)
  \label{expansion2}
\end{equation}
and making an expansion around the saddle point solution we get, for the dominant contribution
to the second term of the right hand side of (\ref{menoscurioso}) in the large lattice volume
limit,

\begin{equation}
\frac{\int\mathrm{d}\theta
  \left<P_u\left(0\right)\right>^2_{\theta} e^{-V E\left(\beta,m,\theta\right)}}
     {\int\mathrm{d}\theta e^{-V E\left(\beta,m,\theta\right)}} =
     \frac{1}{V} \frac{\left(m_u \chi_\eta + \left<S_u\right>\right)^2}{\chi_T}.
  \label{expansion3}
\end{equation}
Since the topological susceptibility $\chi_T$ is linear in $m_u$, for small fermion mass 
(\ref{topsus}), and the scalar condensate $\left<S_u\right>$ is finite in the chiral limit, this
contribution is singular at $m_u=0$.

Equations (\ref{menoscurioso}) and (\ref{expansion3}) show that indeed the pseudoscalar
correlation function in the zero-charge topological sector converges, in the infinite volume
limit, to the pseudoscalar correlation function in the full theory at $\theta=0$. These equations
also show what we can call a \textit{cluster violation} at finite volume for the pseudoscalar
correlation function, in the $Q=0$ topological sector, which disappears in the infinite volume
limit. This \textit{cluster violation} at finite volume is therefore
irrelevant in what concerns the
pseudoscalar correlation function but, conversely, it plays a fundamental role when computing
the pseudoscalar susceptibility in the $Q=0$ topological sector. In fact, if we sum up in
equation (\ref{menoscurioso}) over all lattice points, and take the infinite volume limit,
just in this order, we get for the pseudoscalar susceptibility in the $Q=0$ topological sector

\begin{equation}
\chi^{Q=0}_\eta = \chi_\eta + 
     \frac{\left(m_u \chi_\eta + \left<S_u\right>\right)^2}{\chi_T}.
  \label{expansion4}
\end{equation}
This equation shows that the pseudoscalar susceptibility, in the $Q=0$ sector, diverges in the
chiral limit due to the finite contribution of (\ref{expansion3}) to this susceptibility. 
Hence we have shown that, although the $Q=0$ topological sector breaks spontaneously the
$U_A(1)$ axial symmetry to give account of the anomaly, the Goldstone theorem is not fulfilled
because the divergence of the pseudoscalar susceptibility in this sector does not come from a
divergent correlation length.

The multi-flavor model with spontaneous non-abelian chiral symmetry breaking, as for instance
$QCD$ in the low temperature phase, shows some important differences with respect to the
one-flavor case. The model also suffers from the chiral anomaly, and has a spontaneously broken
$SU_A(N_f)$ chiral symmetry. Because of the Goldstone theorem, there are $N^2_f-1$ massless
pseudoscalar bosons in the chiral limit and, in contrast to the one-flavor case, the
infinite-volume limit and the chiral limit do not commute. However, in what concerns the
flavor-singlet pseudoscalar susceptibility, the essential features previously described for
the one-flavor model still work in the several-flavors case.

Let us consider the simplest case of two degenerate flavors, $m_u=m_d=m$. The anomalous
W-T identity (\ref{Trans-II}) for the flavor-singlet pseudoscalar susceptibility reads
now

\begin{equation}
  \chi_\eta = -\frac{\left<S\right>}{m} -
  \frac{4\chi_T}{m^2}
  \label{Trans-II2f}
\end{equation}
while the non-anomalous identity for the pion susceptibility is

\begin{equation}
  \chi_{\bf \pi} = -\frac{\left<S\right>}{m}
  \label{Trans-pi}
\end{equation}
where $S=S_u+S_d$. The $Q=0$ sector breaks spontaneously the $U_A(2)$ symmetry, and the
W-T identities for this sector are

\begin{equation}
  \chi^{Q=0}_\eta = \chi^{Q=0}_{\bf \pi} = -\frac{\left<S\right>^{Q=0}}{m} -
  \label{Trans-II2fq02f}
\end{equation}

The analysis done in this subsection allows to conclude that, although $\chi_{\bf \pi}$ is a
non-local operator, it takes, in the infinite lattice volume limit, the same value in the
$Q=0$ sector as in the full theory. Conversely, that is not true for the flavor-singlet
pseudoscalar susceptibility, which diverges in the chiral limit in the $Q=0$ sector, 
while remaining finite in the full theory. A straightforward analysis, as the one done for the
one-flavor case, shows that, again, the divergence of $\chi^{Q=0}_\eta$ does not
come from a divergent correlation length.

The case in which the $SU(N_f)$ chiral symmetry is fulfilled in the vacuum will be discussed in
detail in the next subsections.

\subsection{Two flavors and exact $SU_A(2)$ chiral symmetry} \label{Two flavours}

There are several relevant physical theories, as for instance the two-flavor Schwinger model or
$QCD$ in the high temperature phase, that suffer from the $U_A(1)$ axial anomaly, and in which
the non-abelian chiral symmetry is fulfilled in the vacuum. I will discuss in what follows what
are the physical expectations in these theories. 
I will argue that a theory which
verifies the aforementioned properties should show, in the chiral limit, a divergent correlation
length, and a rich spectrum of massless chiral bosons. To this end we will start with the
assumption that all correlation lengths are finite and will see that, in such a case, the
axial $U_A(1)$ symmetry is effectively restored.

We consider a fermion-gauge model with two flavors, up and down, with masses $m_u$ and $m_d$, 
exact $SU_A(2)$ chiral symmetry, and global $U_A(1)$ axial anomaly. The Euclidean fermion-gauge
action (\ref{famnf}) is 

\begin{equation}
  S_F = m_u\bar\psi_u \left(1-\frac{1}{2}D\right)\psi_u +
        m_d\bar\psi_d \left(1-\frac{1}{2}D\right)\psi_d +
        \bar\psi_u D\psi_u + \bar\psi_d D\psi_d
  \label{conlag}
\end{equation}
where $D$ is the Dirac-Ginsparg-Wilson operator.

If we assume, as in the one-flavor model, that all correlation lengths are finite, and the model
shows a mass gap in the chiral limit, the vacuum energy density can also be expanded, as in the
one-flavor case, in powers of the fermion masses $m_u, m_d$, as an ordinary Taylor series

\begin{equation}
  E\left(\beta, m_u, m_d\right) = E\left(\beta, 0, 0\right) -  
  \frac{1}{2} m_u^2 \chi_{s_{u, u}}\left(\beta\right) -
  \frac{1}{2} m_d^2 \chi_{s_{d, d}}\left(\beta\right) -
  m_u m_d \chi_{s_{u, d}}\left(\beta\right) + \dots
\label{taylor2f}
\end{equation}
The linear terms in (\ref{taylor2f}) vanish because the $SU_A(2)$ symmetry is fulfilled
in the vacuum, and 
$\chi_{s_{u, u}}, \chi_{s_{d, d}}$ and $\chi_{s_{u, d}}$ are the scalar up, down and up-down
susceptibilities respectively

$$
\chi_{s_{u, u}}\left(\beta\right) = V \left<S_u^2\right>_{m_u=m_d=0}
$$
$$
\chi_{s_{d, d}}\left(\beta\right) = V \left<S_d^2\right>_{m_u=m_d=0}
$$  
\begin{equation}
\chi_{s_{u, d}}\left(\beta\right) = V \left<S_u S_d\right>_{m_u=m_d=0}
\label{ududs}
\end{equation}
where $S_u$ and $S_d$ are the scalar up and down condensates (\ref{sapcud}), 
normalized by the lattice volume. 
The disconnected contributions are absent in (\ref{ududs}) because the $SU_A(2)$ chiral symmetry
constrains $\left<S_u\right>_{m_u=m_d=0}$ and $\left<S_d\right>_{m_u=m_d=0}$ to vanish, and 
$\chi_{s_{u, u}}\left(\beta\right) = \chi_{s_{d, d}}\left(\beta\right)$ because of flavor
symmetry. Moreover we know that the vacuum energy density of the $Q=0$ topological sector, in the
infinite volume limit, will also be given by (\ref{taylor2f}).\footnote{In Ref. 
  \cite{catorce} it was implicitly assumed that the vacuum energy density of the $Q=0$ sector
  is also a $C^2$ function of $m_u$ and $m_d$. We will show here that this assumption is
  justified.}

In the presence of a $\theta$-vacuum term, expansion (\ref{taylor2f}) becomes 

\begin{equation}
  E\left(\beta, m_u, m_d\right) = E\left(\beta, 0, 0\right) -  
  \frac{1}{2} m_u^2 \chi_{s_{u, u}}\left(\beta\right) -
  \frac{1}{2} m_d^2 \chi_{s_{d, d}}\left(\beta\right) -
  m_u m_d \cos\theta \chi_{s_{u, d}}\left(\beta\right) + \dots
\label{taylor2ftheta}
\end{equation}

The scalar up and down susceptibilities for massless fermions get all their contribution from
the $Q=0$ topological sector, and therefore we can write

$$
\chi_{s_{u, u}}\left(\beta\right) = \chi_{s_{d, d}}\left(\beta\right) = 
\chi_{s_{u, u}}^{Q=0}\left(\beta\right) = \chi_{s_{d, d}}^{Q=0}\left(\beta\right)
$$

Since the $SU_A(2)$ chiral symmetry is fulfilled in the vacuum, the vacuum expectation value of
any local order parameter for this symmetry vanishes in the chiral limit. We have also seen
that any local operator takes, in the thermodynamic limit, the same vacuum expectation value
in the $Q=0$ topological sector than in the full theory. Therefore the $SU(2)_A$ chiral symmetry
of the $Q=0$ sector should also be fulfilled in the vacuum of this sector.

The scalar up and down susceptibilities, in the $Q=0$ sector, for non vanishing quark masses,
also agree with their corresponding values in the full theory, in the infinite volume
limit\footnote{The simplest way to see that is true is to take into account that these
  susceptibilities can be obtained as second-order mass derivatives of the free energy density,
  and the free energy density of the $Q=0$ sector and of the full theory agree if both quark
  masses are of the same sign.}

$$
\chi_{s_{u, u}}^{Q=0}\left(\beta, m_u, m_d\right) =
\chi_{s_{u, u}}\left(\beta, m_u, m_d\right) \hskip 0,5cm
\chi_{s_{d, d}}^{Q=0}\left(\beta, m_u, m_d\right) =
\chi_{s_{d, d}}\left(\beta, m_u, m_d\right).
$$
Therefore these quantities can be obtained from (\ref{taylor2f})

$$
\chi_{s_{u, u}}^{Q=0}\left(\beta, m_u, m_d\right) = \chi_{s_{u, u}}\left(\beta\right) + \dots
$$
$$
\chi_{s_{d, d}}^{Q=0}\left(\beta, m_u, m_d\right) = \chi_{s_{d, d}}\left(\beta\right) + \dots
$$
where the dots indicate terms that vanish in the chiral limit.

The pseudoscalar up and down susceptibilities, in the $Q=0$ sector,
$\chi_{p_{u, u}}^{Q=0}\left(\beta, m_u, m_d\right) = V \left<P_u^2\right>^{Q=0}$,
$\chi_{p_{d, d}}^{Q=0}\left(\beta, m_u, m_d\right) = V \left<P_d^2\right>^{Q=0}$,
can be obtained from the W-T identities in this sector (\ref{wardq0}) beside (\ref{taylor2f})

$$
\chi_{p_{u, u}}^{Q=0}\left(\beta, m_u, m_d\right) = \chi_{s_{u, u}}\left(\beta\right) + 
\frac{\mid m_d\mid}{\mid m_u\mid} \chi_{s_{u, d}}\left(\beta\right) 
+ \dots
$$
\begin{equation}
\chi_{p_{d, d}}^{Q=0}\left(\beta, m_u, m_d\right) = \chi_{s_{d, d}}\left(\beta\right) + 
\frac{\mid m_u\mid}{\mid m_d\mid} \chi_{s_{u, d}}\left(\beta\right) 
+ \dots
\label{fsisb}
\end{equation}
where the absolute value of the quark masses is due to the fact that these susceptibilities are
even functions of the quark masses, and again the dots indicate terms that vanish in the chiral
limit.

The difference of the scalar and pseudoscalar susceptibilities for the up or down quarks, 
$\chi_{s_{u, u}}- \chi_{p_{u, u}}$, $\chi_{s_{d, d}}- \chi_{p_{d, d}}$, is an order parameter
for both, the $U_A(1)$ axial symmetry, and the $SU_A(2)$ chiral symmetry. We can compute this
quantity, in the full theory, making use of (\ref{taylor2f}), the W-T identities
(\ref{Trans-II}), and the topological susceptibility

$$
\chi_T\left(\beta, m_u, m_d\right) = m_u m_d \chi_{s_{u, d}}\left(\beta\right) + \dots
$$
the last obtained from (\ref{taylor2ftheta}), and we get

$$
\chi_{p_{u, u}}\left(\beta, m_u, m_d\right) = -\frac{\left<S_u\right>}{m_u} -
\frac{\chi_T}{m_u^2} = \chi_{s_{u, u}}\left(\beta\right) + \dots
$$
\begin{equation}
\chi_{p_{d, d}}\left(\beta, m_u, m_d\right) = -\frac{\left<S_d\right>}{m_d} -
\frac{\chi_T}{m_d^2} = \chi_{s_{d, d}}\left(\beta\right) + \dots
\label{chisumchipuft}
\end{equation}
where, also in this case, the dots indicate terms that vanish in the chiral limit.
We see from equation (\ref{chisumchipuft}) that indeed, and in spite of the $U_A(1)$
anomaly, $\chi_{s_{u, u}}- \chi_{p_{u, u}}$ and $\chi_{s_{d, d}}- \chi_{p_{d, d}}$,
which are also order parameters for the $SU_A(2)$ chiral symmetry, vanish in the chiral limit,
as it should be.\footnote{A non-local order parameter for a given symmetry, which is fulfilled
in the vacuum, can diverge if the correlation length diverges, as for instance 
the non-linear susceptibility  
$\chi_{nl}\left(h\right) = \pdv [2]{m\left(h\right)}{h}$ in the Ising model at the 
critical temperature. However we are assuming here that all correlation lengths are finite, 
and hence the non-local order parameter should vanish.}

Conversely if we compute this order parameter in the $Q=0$ topological sector we get

$$
\chi_{s_{u, u}}^{Q=0}\left(\beta, m_u, m_d\right) - 
\chi_{p_{u, u}}^{Q=0}\left(\beta, m_u, m_d\right) =
- \frac{\mid m_d\mid}{\mid m_u\mid} \chi_{s_{u, d}}\left(\beta\right) + \dots
$$
$$
\chi_{s_{d, d}}^{Q=0}\left(\beta, m_u, m_d\right) - 
\chi_{p_{d, d}}^{Q=0}\left(\beta, m_u, m_d\right) =
- \frac{\mid m_u\mid}{\mid m_d\mid} \chi_{s_{u, d}}\left(\beta\right) + \dots
$$  
and therefore this order parameter for the non-abelian chiral symmetry only vanishes in the
chiral limit if $\chi_{s_{u, d}}\left(\beta\right) = 0$. Thus we see that, under the assumption
that all correlation lengths are finite, an exact $SU_A(2)$
chiral symmetry in the $Q=0$ sector requires a $\theta$-independent 
vacuum energy density (\ref{taylor2ftheta}), which implies, among other things, that the axial
susceptibility $\chi_{\pi} - \chi_{\delta}$, an order parameter that has been used to test the
effectiveness of the $U_A(1)$ anomaly, vanishes in the chiral limit.

Note, on the other hand, that a non-vanishing value of $\chi_{s_{u, d}}\left(\beta\right)$ not
only implies that the $SU_A(2)$ chiral symmetry of the $Q=0$ sector is spontaneously broken,
but also the $SU_V(2)$ flavor symmetry, 
as follows from (\ref{fsisb}). Even more, a simple calculation of the sum of the flavor singlet
scalar $\chi_\sigma$ and pseudoscalar $\chi_\eta$ susceptibilities for massless quarks give us

\begin{equation}
\chi_{\sigma_{m_u=m_d=0}}^{Q=0} + \chi_{\eta_{m_u=m_d=0}}^{Q=0} =
2 \chi_{s_{u, u}}\left(\beta\right) + 2 \chi_{s_{d, d}}\left(\beta\right) =
4 \chi_{s_{u, u}}\left(\beta\right)
\label{chspchieq0d}
\end{equation}
while if, according to standard Statistical Mechanics, we decompose our degenerate vacuum, or
Gibbs state, into the sum of pure states \cite{parisi} , and calculate $\chi_\sigma + \chi_\eta$
in each one of these pure states, we get

\begin{equation}
\chi_{\sigma_{m_u=m_d=0}}^{Q=0} + \chi_{\eta_{m_u=m_d=0}}^{Q=0} =
4 \chi_{s_{u, u}}\left(\beta\right) +
\frac{1 + \lambda^2}{2 \lambda} \chi_{s_{u, d}}\left(\beta\right)
\label{chspchieq0gs}
\end{equation}
with $\lambda = \frac{\mid m_d\mid}{\mid m_u\mid}$. We see that the consistency between equations
(\ref{chspchieq0d}) and (\ref{chspchieq0gs}) requires again that
$\chi_{s_{u, d}}\left(\beta\right) = 0$.

Therefore, even if one accepts that the $Q=0$ sector spontaneously breaks the $SU_A(2)$ axial and
$SU_V(2)$ flavor symmetries, even though all local order parameters for these symmetries vanish,
we have found that the consistency of the vacuum structure with the theoretical prejudices
about the 
Gibbs state of a statistical system requires, once more, that 
$\chi_{s_{u, u}}\left(\beta\right) = 0$, 
and hence a $\theta$-independent vacuum energy density in the full theory.

In the one-flavor model we have not found inconsistencies between the assumption that the
correlation length is finite, and the physics of the $Q=0$ topological sector. The chiral
condensate takes a non vanishing value in the chiral limit, and hence the $U_A(1)$
axial symmetry is spontaneously broken in the $Q=0$ sector, giving account in this way of the
$U_A(1)$ axial anomaly of the full theory. In the two-flavor model, and under the same
assumption of finiteness of the correlation length, we would need a non-vanishing value of
$\chi_{s_{u,d}}\left(\beta\right)$ to have an effective $U_A(1)$ axial symmetry breaking which,
also in this case, would imply the spontaneous breaking of the global $U_A(1)$ symmetry in the
$Q=0$ sector. 
However, in such a case, we find strong inconsistencies that lead us to conclude that, 
either $\chi_{s_{u,d}}\left(\beta\right) = 0$, and hence the $U_A(1)$ symmetry is effectively
restored, or a divergent correlation length is imperative if the $U_A(1)$ symmetry is not
effectively restored.

\subsection{Landau approach} \label{The phase}

We have argued that the two-flavor theory with exact $SU_A(2)$ chiral symmetry and axial
$U_A(1)$ symmetry violation should exhibit a divergent correlation length in the scalar sector, 
in the chiral limit. In this subsection we will give a qualitative but powerful argument 
which strongly supports this result.
To this end we will explore the expected phase diagram of the model in the $Q=0$ topological
sector \cite{trecebis}, and will apply the Landau theory of phase transitions to it.

Since the $SU_A(2)$ chiral symmetry is assumed to be fulfilled in the vacuum, and the flavor
singlet scalar condensate is an order parameter for this symmetry, its vacuum expectation value 
$\left<S\right>=\left<S_u\right>+\left<S_d\right>=0$ vanishes in the limit in which the fermion
mass $m\rightarrow 0$.
However, if we consider two non-degenerate fermion flavors, up and down, with masses $m_u$ and
$m_d$ respectively, and take the limit $m_u\rightarrow 0$ keeping $m_d\ne 0$ fixed, the up
condensate $S_u$ will reach a non-vanishing value

\begin{equation}
  \lim_{m_u\rightarrow 0} \left<S_u\right> = s_u\left(m_d\right) \ne 0
\label{ucond}
\end{equation}
because the $U(1)_u$ axial symmetry, which exhibits our model when $m_u=0$, is anomalous, and the
$SU_A(2)$ chiral symmetry, which would enforce the up condensate to be zero, is explicitly
broken if $m_d\ne 0$.

Obviously the same argument applies if we interchange $m_u$ and $m_d$, and we can therefore write

\begin{equation}
  \lim_{m_d\rightarrow 0} \left<S_d\right> = s_d\left(m_u\right) \ne 0
\label{dcond}
\end{equation}
and since the $SU_A(2)$ chiral symmetry is recovered and fulfilled in
the vacuum when $m_u, m_d\rightarrow 0$, we get

\begin{equation}
\lim_{m_d\rightarrow 0} s_u\left(m_d\right) = \lim_{m_u\rightarrow 0} s_d\left(m_u\right) = 0
\label{udcond}
\end{equation}

Let us consider now our model, with two non degenerate fermion flavors, restricted to the
$Q=0$ topological sector. As previously discussed, 
the mean value of any local or intensive operator in 
the $Q=0$ topological sector will be equal, if we restrict ourselves to the region in which
both $m_u>0$, and $m_d>0$, to its mean value in the full theory, in the infinite
volume limit.\footnote{Since the two flavor model with $m_u<0$ and $m_d<0$ at
$\theta=0$ is equivalent to the same model with $m_u>0$ and $m_d>0$, this result is also true
if both $m_u<0$ and $m_d<0$.} We can hence apply this result to $\left<S_u\right>$ and
$\left<S_d\right>$ and write the following equations

$$\lim_{m_u\rightarrow 0} \left<S_u\right>^{Q=0} = s_u\left(m_d\right) \ne 0$$
\begin{equation}
  \lim_{m_d\rightarrow 0} \left<S_d\right>^{Q=0} = s_d\left(m_u\right) \ne 0
\label{dcondq0}
\end{equation}
The global $U(1)_u$ axial symmetry of our model at $m_u=0$, and the $U(1)_d$
symmetry at $m_d=0$, are not anomalous in the $Q=0$ sector, and equation (\ref{dcondq0}) tells
us that both, the $U(1)_u$ symmetry at $m_u=0, m_d\ne 0$ and the $U(1)_d$ symmetry at
$m_u\ne0, m_d=0$ are spontaneously broken in this sector. This is not surprising at all since
the present situation is similar to what happens in the one flavor model previously discussed.

\begin{figure}[h!]
  \centerline{\includegraphics[scale=1.03]{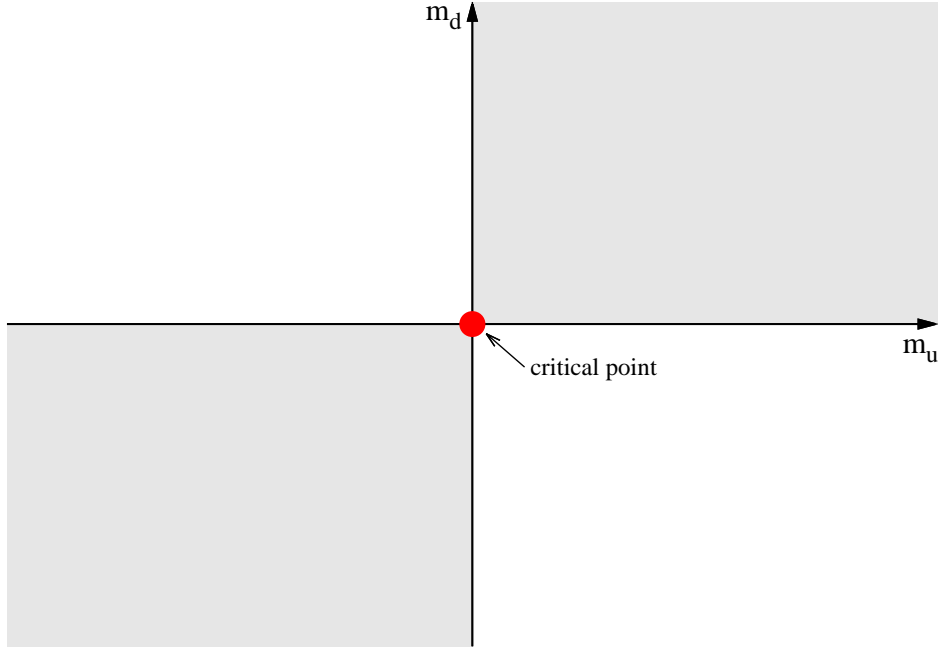}}
  \caption{Phase diagram of the two-flavor model in the $Q=0$ topological sector. The coordinate 
axis in the $(m_u,m_d)$ plane are first order phase transition lines. The origin of coordinates 
is the end point of all first order transition lines. The
vacuum energy density, its derivatives, and expectation values of local operators of
the two-flavor model at $\theta=0$ only agree with those of the $Q=0$ sector in the first
$(m_u>0,m_d>0)$ and third $(m_u<0,m_d<0)$ quadrants (the darkened areas).}
\end{figure}

Fig. 1 is a schematic representation of the phase diagram of the two-flavor model, in the
$Q=0$ topological sector, and in the $(m_u, m_d)$ plane, which emerges from the previous
discussion. The two coordinate axis show first order phase transition lines. If we cross
perpendicularly the $m_d=0$ axis, the mean value of the down condensate jumps from $s_d(m_u)$
to $-s_d(m_u)$, and the same is true if we interchange up and down. All first order
transition lines end however at a common point, the origin of coordinates $m_u=m_d=0$,
where all condensates vanish because at this point we recover the $SU_A(2)$ chiral symmetry,
which is assumed to be also a symmetry of the vacuum. Notice that if the $SU_A(2)$ chiral
symmetry is spontaneously broken, as it happens for instance in the low temperature phase
of $QCD$, the phase diagram in the $(m_u, m_d)$ plane would be the same as that of Fig. 1
with the only exception that the origin of coordinates is not an end point.

Landau's theory of phase transitions predicts that the end point placed at the origin of
coordinates in the $(m_u, m_d)$ plane is a critical point, the scalar condensate should show
a non analytic dependence on the fermion masses $m_u$ and $m_d$ as we approach the critical
point, and hence the scalar susceptibility should diverge. But since the vacuum energy
density in the $Q=0$ topological sector, and its fermion mass derivatives, matches the vacuum
energy density and fermion mass derivatives in the full theory,
and the same is true for the critical equation of state, Landau's theory
of phase transitions
predicts a non-analytic dependence of the flavor singlet scalar condensate on the
fermion mass, and a divergent correlation length in the chiral limit of our full theory, in which
we take into account the contribution of all topological sectors.

More precisely, we can apply the Landau approach to analyze the critical behavior around the two
first order transition lines in Fig.1 near the end point, or critical point. In the analysis of the
$m_d=0$ transition line we consider $m_d$ as an external ''magnetic field''and $m_u$ as the
''temperature'', and vice versa for the analysis of the $m_u=0$ line. Then the standard Landau
approach tell us that the up and down condensates verify the two following equations of state

$$
-m_u\left<S_u\right>^{-3} = -2 C_1 m_d \left<S_u\right>^{-2} + 4 C_2
$$

\begin{equation}
-m_d\left<S_d\right>^{-3} = -2 C_1 m_u \left<S_d\right>^{-2} + 4 C_2
\label{eosud}
\end{equation}
where $C_1$ and $C_2$ are two positive constants. If we fix the ratio of the up and down masses
$\frac{m_u}{m_d}=\lambda$, the equations of state (\ref{eosud}) allow us to write the following
expansions for de up and down condensates

$$
  \left<S_u\right> = -m_u^{\frac{1}{3}}\left(\left(\frac{1}{4C_2}\right)^\frac{1}{3} +
\frac{C_1}{3\left(2C_2^2\right)^\frac{1}{3}\lambda} m_u^\frac{1}{3} + \ldots\right)
$$
\begin{equation}
  \left<S_d\right> = -m_d^{\frac{1}{3}}\left(\left(\frac{1}{4C_2}\right)^\frac{1}{3} +
\frac{C_1\lambda}{3\left(2C_2^2\right)^\frac{1}{3}} m_d^\frac{1}{3} + \ldots\right)
\label{expud}
\end{equation}
Equation
(\ref{expud}) shows explicitly the non analytical behavior of the up and down condensates.
In the degenerate flavor case, $m_u=m_d=m$, the scalar condensate and the flavor-singlet scalar
susceptibility near the critical point scale as

$$
\left<S\right> = \left<S_u\right> + \left<S_d\right> = 
-\left(\frac{2}{C_2}\right)^\frac{1}{3} m^\frac{1}{3} + \ldots
$$
\begin{equation}
  \chi_\sigma\left(m\right) =  \frac{1}{3} \left(\frac{2}{C_2}\right)^\frac{1}{3} m^{-\frac{2}{3}}
  + \ldots
\label{landaucond}
\end{equation}
showing up explicitly the divergence of the flavor singlet scalar susceptibility in the
chiral limit.

We see that the critical behavior of the chiral condensate in the Landau approach
(\ref{landaucond}) is described by the mean field critical exponent $\delta=3$.  
Mean field critical exponents are expected to be correct in high dimensions, while in low
dimensions, the effect of fluctuations can change their mean field values. This means that, 
in the latter case, the Landau approach give us a good qualitative description of the phase
diagram, but fails in its quantitative predictions of critical exponents.

To finish the Landau approach analysis we want to point out that all these results can be
generalized in a straightforward way to a number of flavors $N_f>2$.

\subsection{Critical behavior of the two flavor model with an isospin breaking term}
\label{cbtf}

Beyond the Landau approach, we can 
parameterize the critical behavior of the flavor singlet scalar
condensate and of the mass-dependent contribution to the vacuum energy density, in the
two degenerate flavor model, with a critical exponent $\delta>1$

\begin{equation}
  \left<S\right>_{m\rightarrow 0} \simeq -C m^{\frac{1}{\delta}}.
\label{criticaldelta}
\end{equation}

\begin{equation}
  E\left(\beta, m\right) - E\left(\beta, 0\right)\simeq -\frac{C\delta}{\delta+1}
  m^{\frac{\delta+1}{\delta}}.
\label{criticalved}
\end{equation}
where $C$ is a dimensionless positive constant that depends on the inverse gauge coupling $\beta$.
Equation (\ref{criticaldelta}) gives us a divergent scalar susceptibility, 
$\chi_\sigma\left(m\right)\sim \frac{C}{\delta} m^\frac{1-\delta}{\delta}$, and hence a
massless scalar boson as $m\rightarrow 0$. 

If on the other hand we write the W-T identity for the isotriplet of ''pions''
which follows from the $SU_A(2)$ non-anomalous chiral symmetry 

\begin{equation}
\chi_{\bar\pi}\left(m\right) = -\frac{\left\langle S\right\rangle}{m},
\label{chipi2}
\end{equation}
we get that also $\chi_{\bar\pi}\left(m\right)$ diverges when $m\rightarrow 0$ as
$C m^\frac{1-\delta}{\delta}$, and a rich
spectrum of massless bosons $(\sigma, \bar\pi)$ emerges in the chiral limit.
The susceptibility of the flavor singlet pseudoscalar condensate fulfills the anomalous
W-T identity (\ref{Trans-II2f}), and because of the $U_A(1)$ axial anomaly, 
the $\eta$-boson mass is expected to remain finite (non-vanishing) in the chiral limit.

The hyperscaling hypothesis, which arises as a natural consequence of the block-spin
renormalization group approach, says that the only relevant length near the critical point of
a \textit{magnetic}
system, in what concerns the singular part $E_s\left(\beta, m\right)$ of the free or vacuum
energy density, 
is the correlation length $\xi$. Since equation (\ref{criticalved}) contains only the singular
contribution to the vacuum energy density, we can write

\begin{equation}
  E_s\left(\beta, m\right) \simeq -\frac{C\delta}{\delta+1} m^{\frac{\delta+1}{\delta}}
  \sim \xi^{-d}
\label{criticalvedhyp}
\end{equation}
and the following relationship between the correlation length and the fermion mass 

\begin{equation}
  \xi \sim m^{-\frac{\delta+1}{d\delta}}
\label{hyplcm}
\end{equation}
which implies that the pion and sigma-meson masses scale with the fermion mass as follows

\begin{equation}
  m_{\bar\pi}, m_\sigma \sim m^{\frac{\delta+1}{d\delta}}
\label{hyplcmp}
\end{equation}

In the presence of an isospin breaking term, the fermion action can be written in a compact form as

\begin{equation}
  S_F = \left(\frac{m_u+m_d}{2}\right)\bar\psi \left(1-\frac{1}{2}D\right)\psi -
  \left(\frac{m_d-m_u}{2}\right)\bar\psi \left(1-\frac{1}{2}D\right)\tau_3\psi +
  \bar\psi D\psi
  \label{conlagcomp}
\end{equation}
where $\psi$ is a Grassmann field carrying site,
Dirac, color and flavor indices, and $\tau_3$ is the third Pauli matrix acting in flavor space.

If we include also a $\theta$-vacuum term in the action, this $\theta$-term can be removed
through a chiral $U_A(1)$ transformation, which leaves the $\bar\psi D\psi$ interaction
term invariant, and if next we also perform a suitable non-anomalous chiral transformation,
we get the effective fermion action that follows

$$
  S_F = M\left(m_u,m_d,\theta\right)\bar\psi \left(1-\frac{1}{2}D\right)\psi +
  A\left(m_u,m_d,\theta\right)i\bar\psi\gamma_5 \left(1-\frac{1}{2}D\right)\psi
$$
\begin{equation}
  \hskip -3,4cm
  + B\left(m_u,m_d,\theta\right)\bar\psi \left(1-\frac{1}{2}D\right)\tau_3\psi + 
  \bar\psi D\psi
  \label{efa2f}
\end{equation}
where $M\left(m_u,m_d,\theta\right)$, $A\left(m_u,m_d,\theta\right)$ and
$B\left(m_u,m_d,\theta\right)$ are given by

\begin{equation}
  M\left(m_u,m_d,\theta\right) = \frac{1}{2} \left(m^2_u + m^2_d +
  2 m_u m_d \cos\theta\right)^{\frac{1}{2}}
  \label{M}
\end{equation}

\begin{equation}
  A\left(m_u,m_d,\theta\right) = 
  \frac{2 m_u m_d\sin\frac{\theta}{2}}
  {\left(m_u + m_d\right)\left(1 + \frac{m^2_u + m^2_d - 2 m_u m_d}{m^2_u + m^2_d + 2 m_u m_d}
    \tan^2\frac{\theta}{2}\right)^\frac{1}{2}}
    \label{A}
\end{equation}

\begin{equation}
  B\left(m_u,m_d,\theta\right) = -\frac{m_d - m_u}
  {2 \cos\frac{\theta}{2}\left(1 + \frac{m^2_u + m^2_d - 2 m_u m_d}{m^2_u + m^2_d + 2 m_u m_d}
    \tan^2\frac{\theta}{2}\right)^\frac{1}{2}}
    \label{B}
\end{equation}

Since we do not expect singularities at non-vanishing fermion masses, the vacuum energy density
$E\left(\beta, M, A, B\right)$ can be expanded in powers of $A$ and $B$ as an ordinary Taylor
series, and taking into account the symmetries of the effective action (\ref{efa2f}), we can
write the following equation for this expansion up to second order

\begin{equation}
    E\left(\beta, m_u, m_d, \theta\right)\equiv 
  E\left(\beta, M, A, B\right) = E\left(\beta, M, 0, 0\right) +
  \frac{1}{2} A^2 \chi_\eta \left(\beta, M\right) +
  \frac{1}{2} B^2 \chi_\delta \left(\beta, M\right) + \dots
    \label{vede2o}
\end{equation}
where $\chi_\eta \left(\beta, M\right)$ and $\chi_\delta \left(\beta, M\right)$ 
are the flavor singlet pseudoscalar susceptibility and the $\delta$-meson susceptibility in the
theory with two degenerate flavors of mass $M(m_u, m_d, \theta)$, respectively. Note that this
expansion should have a good convergence if $\theta$ and $m_d-m_u$ are small.

The vacuum energy density, to the lowest order of the expansion (\ref{vede2o}), is that
of the model
with two degenerate flavors of mass $M(m_u, m_d, \theta)$, in the absence of a $\theta$-vacuum
term. We have previously shown that this model should show a critical behavior
(\ref{criticaldelta}), (\ref{criticalved}) around the chiral limit, and hence we get, to the
lowest order of this expansion, 

\begin{equation}
  E\left(\beta, m_u, m_d, \theta\right) - E\left(\beta, 0, 0, 0\right)
  = -\frac{C}{2^{\frac{\delta+1}{\delta}}}\frac{\delta}{\delta+1}
  \left(m^2_u + m^2_d +
      2 m_u m_d \cos\theta\right)^\frac{\delta+1}{2\delta} + \dots
    \label{vede2olo}
\end{equation}
The free energy density depends on $m_u$, $m_d$ and $\theta$ through 
$\left(m^2_u + m^2_d + 2 m_u m_d \cos\theta\right)^\frac{1}{2}$, and its dominant
contribution in the chiral limit is given by the power-law behavior of equation (\ref{vede2olo}).
\footnote{Note that if we apply this expansion of the vacuum energy density to two-flavor $QCD$
  at $T=0$, where chiral symmetry is spontaneously broken, and hence $\delta=\infty$ in
  (\ref{criticaldelta}) and (\ref{criticalved}), we get the vacuum energy density of
  the low energy chiral effective Lagrangian model \cite{Smilga}.}

The flavor-singlet pseudoscalar susceptibility, $\chi_\eta\left(\beta, M\right)$, fulfills the
anomalous W-T identity (\ref{Trans-II2f}), and hence it is expected to remain
finite in the chiral limit. Since the $SU_A(2)$ chiral symmetry is exact in this limit,
the same holds true for $\chi_\delta\left(\beta, M\right)$. In such conditions, the relevance of
the second order correction to the zero-order contribution to the vacuum energy density
(\ref{vede2o}), for two degenerate flavors, turns out to be

\begin{equation}
\frac{A^2\left(m,\theta\right)}{E\left(\beta, M, 0\right)-E\left(\beta, 0, 0\right)}
\sim m^{1-\frac{1}{\delta}}\frac{\sin^2\frac{\theta}{2}}
{\left(\cos\frac{\theta}{2}\right)^{1+\frac{1}{\delta}}}
\label{socveddf}
\end{equation}
while in the isospin breaking case, and for small $\theta$ values, we have

$$
\frac{A^2\left(m_u,m_d,\theta\right)}{E\left(\beta, M, 0, 0\right)-E\left(\beta, 0, 0, 0\right)}
\sim \frac{m^2_u m^2_d \theta^2}{\left(m_u + m_d\right)^{3+\frac{1}{\delta}}}
$$
\begin{equation}
\frac{B^2\left(m_u,m_d,\theta\right)}{E\left(\beta, M, 0, 0\right)-E\left(\beta, 0, 0, 0\right)}
\sim \frac{\left(m_d - m_u\right)^2}{\left(m_u + m_d\right)^{1+\frac{1}{\delta}}}
\label{socved}
\end{equation}

Since $\delta>1$ ($\delta=3$ in the mean field model), we see that the critical behavior
of the model, which describes the low energy theory, is fully controlled in both cases by
the zero-order
contribution to the vacuum energy density (\ref{vede2olodf}), and the second order contribution 
can be neglected in what concerns the chiral limit of the theory.

Let us now look  at some 
interesting physical consequences that can be obtained from equation (\ref{vede2olo}).
In the degenerate flavor case, $m_u=m_d=m$, equations (\ref{vede2olo}), (\ref{hyplcm}) and
(\ref{hyplcmp}) become

\begin{equation}
  E\left(\beta, m, \theta\right) - E\left(\beta, 0, 0\right)= 
  -\frac{C\delta}{\delta+1} \left(m\cos\frac{\theta}{2}\right)^\frac{\delta+1}{\delta} + \dots
  \label{vede2olodf}
\end{equation}

\begin{equation}
  \xi \sim \left(m \cos\frac{\theta}{2}\right)^{-\frac{\delta+1}{d\delta}}
\label{hyplcmdf}
\end{equation}

\begin{equation}
  m_{\bar\pi}, m_\sigma \sim \left(m \cos\frac{\theta}{2}\right)^{\frac{\delta+1}{d\delta}}
\label{hyplcmpdf}
\end{equation}

For non-degenerate flavors, the 
vacuum energy density (\ref{vede2olo}) at $\theta=0$ is a function of $m_u+m_d$, hence the
vacuum expectation values of the up and down condensates are equal, and the same holds true for
their susceptibilities:

$$
\left<S_u\right> = \left<S_u\right> =
-\frac{C}{2^{\frac{\delta+1}{\delta}}}
\left(m_u + m_d\right)^\frac{1}{\delta}
$$
$$
  \sum_x \left(\left<S_u\left(x\right) S_u\left(0\right)\right>
  - \left<S_u\left(x\right)\right>\left<S_u\left(0\right)\right>\right) = 
  \sum_x \left(\left<S_d\left(x\right) S_d\left(0\right)\right>
  - \left<S_d\left(x\right)\right>\left<S_d\left(0\right)\right>\right) =
$$
  \begin{equation}
    \hskip -3cm
    \sum_x \left(\left<S_u\left(x\right) S_d\left(0\right)\right>
  - \left<S_u\left(x\right)\right>\left<S_d\left(0\right)\right>\right) =
  \frac{C}{2^{\frac{\delta+1}{\delta}}} \frac{1}{\delta}
\left(m_u + m_d\right)^\frac{1-\delta}{\delta}.
  \label{qcas}
\end{equation}
We don't see any dependency on $m_d-m_u$, and 
isospin breaking effects are therefore absent in these quantities, which on the other hand
show a singular behavior in the chiral limit. The normalized flavor singlet scalar susceptibility,
$\chi_\sigma$,

  \begin{equation}
    \chi_\sigma =
  \frac{C}{2^{\frac{1}{\delta}}} \frac{1}{\delta}
\left(m_u + m_d\right)^\frac{1-\delta}{\delta}.
  \label{sigmasuso0}
\end{equation}
diverges in the chiral limit, while the $\delta$-meson susceptibility, $\chi_\delta$, 
vanishes in the zero-order
approximation to the vacuum energy density, this indicating that it is a good approximation when
the ratio of the $\sigma$ and $\delta$ meson masses is small, $\frac{m\sigma}{m_\delta}\ll 1$.

The topological susceptibility is given by

  \begin{equation}
    \chi_T =
  \frac{C}{2^{\frac{\delta+1}{\delta}}}
\left(m_u + m_d\right)^\frac{1-\delta}{\delta} m_u m_d
  \label{topsuso0}
\end{equation}
showing that this quantity is sensitive to the isospin breakdown.

The W-T identity for the charged pions, $\pi^\pm$, reads

  \begin{equation}
    \chi_{\pi^\pm} = -\frac{\left<S_u\right> + \left<S_d\right>}{m_u + m_d}
  \label{wtipmm}
\end{equation}
and hence we get

  \begin{equation}
    \chi_{\pi^\pm} = \frac{C}{2^{\frac{1}{\delta}}}
\left(m_u + m_d\right)^\frac{1-\delta}{\delta}.
  \label{pmmso0}
\end{equation}
Like the $\sigma$-susceptibility, the charged pions susceptibility diverges in the chiral limit.
 
To calculate the susceptibility of the neutral pion, we use the following $W-T$ identities

$$
\sum_x \left<P_u\left(x\right) P_u\left(0\right)\right> = - \frac{\left<S_u\right>}{m_u}
- \frac{\chi_T}{m_u^2}
$$
$$
\sum_x \left<P_d\left(x\right) P_d\left(0\right)\right> = - \frac{\left<S_d\right>}{m_d}
- \frac{\chi_T}{m_d^2}
$$
\begin{equation}
\hskip -0,8cm
\sum_x \left<P_u\left(x\right) P_d\left(0\right)\right> = - \frac{\chi_T}{m_u m_d}
    \label{wtipupd}
\end{equation}
which give us 

\begin{equation}
\hskip -0,8cm
\sum_x \left<P_u\left(x\right) P_u\left(0\right)\right> = 
\sum_x \left<P_d\left(x\right) P_d\left(0\right)\right> = 
-\sum_x \left<P_u\left(x\right) P_d\left(0\right)\right> =
\frac{C}{2^{\frac{\delta+1}{\delta}}}
\left(m_u + m_d\right)^\frac{1-\delta}{\delta}
\label{wtipupdr}
\end{equation}
and for the normalized neutral pion susceptibility we get

  \begin{equation}
    \chi_{\pi^0} = \frac{C}{2^{\frac{1}{\delta}}}
\left(m_u + m_d\right)^\frac{1-\delta}{\delta}.
  \label{p0so0}
\end{equation}

Equations (\ref{pmmso0}) and (\ref{p0so0}) show that the $\pi^\pm$ and
$\pi^0$susceptibilities are equal and
independent of $m_d-m_u$. Again isospin breaking effects are absent in these quantities, and
even though $m_d-m_u\ne 0$, the three pions have the same mass. In what concerns the flavor-singlet
pseudoscalar susceptibility, $\chi_\eta$, equation (\ref{wtipupdr}) shows that it vanishes.

Last but not least, if for simplicity we consider two degenerate flavors, equations
(\ref{hyplcmp}) and (\ref{topsuso0}) imply that the pion mass $m_{\bar\pi}$
(or the $\sigma$-meson mass) and the topological susceptibility $\chi_T$ verify the following
relation

  \begin{equation}
    \frac{m_{\bar\pi}}{\left(\chi_T\right)^{\frac{1}{d}}} = k\left(\beta, L_t\right)
  \label{mpichitr}
\end{equation}
where $k$ is a dimensionless quantity that depends on the inverse gauge coupling $\beta$, and
eventually, at finite temperature $T$, on the lattice temporal extent $L_t$, but that is
independent of the fermion mass $m$.

In summary, we have seen that, in the zero order approximation to the vacuum energy
density, that accounts for the chiral critical behavior of the theory, 
isospin breaking effects only manifest in the topological susceptibility. The three pions
have the same mass, the ratio of the pion (\ref{p0so0}) and $\sigma$-meson
(\ref{sigmasuso0}) susceptibilities is equal to the critical exponent $\delta$, 
and the pion (or $\sigma$-meson) mass is related with the topological
susceptibility, as seen in equation (\ref{mpichitr}).

\section{Two-flavor Schwinger model as a test bed} \label{Schwinger}

Quantum Electrodynamics in $(1+1)$-dimensions, is a good laboratory to
test the results reported in the previous section. The model is confining \cite{kogut1},
exactly solvable at zero fermion mass, has non-trivial topology, and 
shows explicitly the $U_A(1)$ axial anomaly \cite{kogut2}. Besides that,
the Schwinger model does not require infinite renormalization, and this means that, if we use a
lattice regularization, the bare parameters remain finite in the continuum limit.

On the other hand, the $SU_A(N_f)$ non-anomalous axial symmetry in the chiral limit of the
multi-flavor Schwinger model is fulfilled
in the vacuum, and this property makes this model a perfect candidate to check
our predictions on the existence of quasi-massless scalar and pseudoscalar bosons in the
spectrum of the model, the mass of which vanishes in the chiral limit. 

The Euclidean continuum action for the two-flavor theory is

$$
  S = \int \mathrm{d}^2x \{\bar\psi_u\left(x\right)
    \gamma_\mu\left(\partial_\mu +
    iA_\mu\left(x\right)\right)\psi_u\left(x\right) +
    \bar\psi_d\left(x\right)
    \gamma_\mu\left(\partial_\mu +
        iA_\mu\left(x\right)\right)\psi_d\left(x\right)\} + 
$$
\begin{equation}
\hskip -0,5cm          
  \int \mathrm{d}^2x \{m_u \bar\psi_u\left(x\right)\psi_u\left(x\right) +
  m_d \bar\psi_d\left(x\right)\psi_d\left(x\right)
  + \frac{1}{4e^2} F^2_{\mu\nu}\left(x\right) + i\theta Q\left(x\right)\}
  \label{dos}
\end{equation}
where $m_u$, $m_d$ are the fermion masses and $e$ is the electric charge or gauge coupling,
which has mass dimensions. $F_{\mu\nu}(x) = \partial_\mu A_\nu(x) - \partial_\nu A_\mu(x)$,
and $\gamma_\mu$ are $2\times 2$ matrices satisfying the algebra

\begin{equation}
  \{\gamma_\mu, \gamma_\nu\} = 2 g_{\mu\nu}
\end{equation}

At the classic level this theory has an internal $SU_V(2)\times SU_A(2)\times U_V(1)\times U_A(1)$
symmetry in the chiral limit. However, the $U_A(1)$-axial symmetry is broken at the quantum level
because of the axial anomaly. The divergence of the axial current is

\begin{equation}
  \partial_\mu J^A_\mu(x) = \frac{1}{2\pi} \epsilon_{\mu\nu} F_{\mu\nu}(x),
  \label{currentdivergence}
\end{equation}

\noindent
where $\epsilon_{\mu\nu}$ is the antisymmetric tensor, and hence does not vanish. 
The axial anomaly induces the topological $\theta$-term $i\theta Q=i\theta \int \mathrm{d}^2x Q(x)$
in the action, where 

\begin{equation}
  Q\left(x\right) = \frac{1}{4\pi} \epsilon_{\mu\nu} F_{\mu\nu}(x)
  \label{schtc}
\end{equation}
is the density of topological charge, the topological charge $Q$ being an integer number.

The Schwinger model was analyzed years ago by Coleman \cite{coleman}, computing some
quantitative properties of the theory in the continuum for both, weak coupling 
$\frac{e}{m}\ll 1$, and strong coupling or chiral limit $\frac{e}{m}\gg 1$.

For the one-flavor case, Coleman computed the particle spectrum of the model, which shows a mass
gap in the chiral limit, and conjectured 
the existence of a phase transition at $\theta=\pi$ and some intermediate fermion mass $m$
separating a weak coupling phase ($\frac{e}{m}\ll 1$), where the $Z_2$ symmetry of the
model at $\theta=\pi$ is spontaneously broken, from a strong coupling phase
($\frac{e}{m}\gg 1$), in which the $Z_2$ symmetry is fulfilled in the vacuum. This qualitative
result has been recently confirmed by numerical simulations of the Euclidean-lattice version of
the model \cite{monos}. 

What is however more interesting for the content of this article is the Coleman analysis of the
two-flavor model. As previously stated, the theory (\ref{dos}) has an internal
$SU_V(2)\times SU_A(2)\times U_V(1)\times U_A(1)$
symmetry in the chiral limit, and the $ U_A(1)$ axial symmetry is anomalous. Since continuous
internal symmetries can not be spontaneously broken in a local field theory in two dimensions
\cite{coleman2}, the $SU_A(2)$ symmetry has to be fulfilled in the
vacuum, and the scalar condensate, which is an order parameter for this symmetry, vanishes in
the chiral limit. Hence the two-flavor
Schwinger model verifies all the conditions we assumed in section \ref{The main}.

We summarize here the main Coleman's findings for the two-flavor model with degenerate masses
$m_u=m_d=m$:

\begin{enumerate}
\item 
  For weak coupling, $\frac{e}{m}\ll 1$, the results on the particle spectrum are almost the
  same as for the massive Schwinger model.  
\item
  For strong coupling, $\frac{e}{m}\gg 1$, the low-energy effective theory depends only on one
  mass parameter, $m^\frac{2}{3} e^\frac{1}{3} \cos^\frac{2}{3}\frac{\theta}{2}$, the vacuum
  energy density is then proportional to

  \begin{equation}
    E\left(m, e, \theta\right) \propto e^\frac{2}{3}
    \left(m\cos\frac{\theta}{2}\right)^\frac{4}{3},
  \label{vacsch}
\end{equation}
and the chiral condensate, at $\theta=0$, is therefore

  \begin{equation}
  \langle \bar\psi\psi\rangle \propto m^\frac{1}{3} e^\frac{2}{3}
  \label{chiralsch}
\end{equation}

\item
  The lightest particle in the theory is an
  isotriplet, and the next lightest is an isosinglet. The
  isosinglet/isotriplet mass ratio is $\sqrt 3$. If there are other
  stable particles in the model, they must be
  $O\left(\left[\frac{e}{m}\right]^\frac{2}{3}\right)$
  times heavier than these. The light boson mass, $M$, has a fractional power dependence on the
  fermion mass $m$:

  \begin{equation}
  M\propto e^\frac{1}{3} \left(m\cos\frac{\theta}{2}\right)^\frac{2}{3}
  \label{colmass}
  \end{equation}

\end{enumerate}

Many of these results have been corroborated by several authors both, in the continuum
\cite{smilga2}, \cite{seiler}, \cite{james}, \cite{jac}, \cite{smilga3}, and using the lattice
approach \cite{gks}, \cite{gattr}. Coleman concluded his paper \cite{coleman} by asking some
questions concerning things he didn't understand, and we cite here two of them:

\begin{enumerate}
\item
  Why are the lightest particles in the theory a degenerate isotriplet, even if one quark is $10$ 
  times heavier than the other?  
\item
  Why does the next-lightest particle has $I^{PG} = 0^{++}$, rather than $0^{--}$?

\end{enumerate}

The results of section \ref{The main} allow us to qualitatively understand the main Coleman's
findings for the two-flavor model with degenerate masses in the strong coupling limit, as well as 
to give a reliable answer to the previous questions.

In section \ref{cbtf} we predicted, from the interplay between the $U_A(1)$ anomaly and the exact
$SU_A(2)$ chiral symmetry, a singular behavior of the vacuum energy density (\ref{criticalved}),
(\ref{vede2olodf}) in the chiral limit limit  as 
$E\sim C \left(m\cos\frac{\theta}{2}\right)^{\frac{\delta+1}{\delta}}$ . In the Schwinger
model, a simple dimensional analysis tell us that $C$ must be proportional to
$e^{\frac{\delta-1}{\delta}}$. Therefore our result matches perfectly Coleman's result
(\ref{vacsch}) if we choose $\delta=3$.

In what concerns the masses of the light bosons, our prediction (\ref{hyplcmpdf}),
$m_{\bar\pi}, m_\sigma \sim \left(m \cos\frac{\theta}{2}\right)^{\frac{\delta+1}{d\delta}}$
matches, for $\delta=3$, Colemans's result (\ref{colmass}) too.

In section \ref{cbtf} we also predicted that the flavor-singlet scalar susceptibility 
(\ref{sigmasuso0}), and the ''pion'' susceptibility (\ref{pmmso0}), (\ref{p0so0}), 
should diverge in the chiral 
limit as $\frac{K}{\delta} m^\frac{1-\delta}{\delta} e^\frac{\delta-1}{\delta}$ 
and $K m^\frac{1-\delta}{\delta} e^\frac{\delta-1}{\delta}$ respectively,\footnote{The factor
$e^\frac{\delta-1}{\delta}$ comes again from dimensional analysis in the Schwinger model.}
and for $\delta=3$ we have

\begin{equation}
  {\chi_\sigma}_{m\rightarrow 0} = \frac{K}{3} m^{-\frac{2}{3}} e^{\frac{2}{3}} 
  \sim \frac{\mid\langle 0\mid \hat{O}_{\sigma}\mid
    \sigma\rangle\mid^2}{m_\sigma} \hskip 1cm
       {\chi_{\pi^0}}_{m\rightarrow 0} = K m^{-\frac{2}{3}}e^{\frac{2}{3}} 
       \sim \frac{\mid\langle 0\mid \hat{O}_{\pi^0}\mid \pi^0
      \rangle\mid^2}
    {m_{\pi^0}}
\label{lasdosck}
\end{equation}
where $K$ is a dimensionless constant.

We have also seen that the $\sigma$ and $\bar\pi$ meson masses, in the strong-coupling limit, scale
with the quark mass as 

\begin{equation}
m_{\bar\pi}, m_\sigma \sim m^{\frac{2}{3}} e^{\frac{1}{3}}.
\label{mpimsimq}
\end{equation}
Taking into account that the $SU_A(2)$ symmetry is exact in the chiral limit, equations
(\ref{lasdosck}) and (\ref{mpimsimq}) imply that

\begin{equation}
  \lim_{m\rightarrow 0}\mid\langle 0\mid \hat{O}_{\sigma}\mid\sigma\rangle\mid^2 =
      \lim_{m\rightarrow 0}\mid\langle 0\mid \hat{O}_{\pi^0}\mid \pi^0\rangle\mid^2
      \sim e
\label{elmaigu}
\end{equation}
and therefore we have

\begin{equation}
\lim_{m\rightarrow 0} \frac{\chi_{\pi^0}\left(m, e\right)}{\chi_\sigma\left(m, e\right)} =
\lim_{m\rightarrow 0} \frac{m_\sigma\left(m, e\right)}{m_{\pi^0}\left(m, e\right)} = 3
\label{msigmpies3}
\end{equation}

These results show that indeed the lightest particle in the theory is an
isotriplet, and the next lightest is an isosinglet $I^{PG} = 0^{++}$. However our result for the
ratio $\frac{m_\sigma}{m_{\bar\pi}} = 3$ \cite{catorce} is in disagreement with Coleman's result 
$\frac{m_\sigma}{m_{\bar\pi}} = \sqrt 3$ \cite{coleman}.

In what concerns the first Coleman's question, we have argued in section \ref{cbtf} that the
strong coupling limit performed by Coleman corresponds to the zero-order contribution to
the vacuum energy density expansion (\ref{vede2o}). This zero-order contribution depends
on the quark masses only through the combination $m_u+m_d$, and we have shown that in such a case
only the topological susceptibility is sensitive to isospin breaking effects. The three pion
susceptibilities (\ref{pmmso0}), (\ref{p0so0}) and masses are equal, and to see isospin breaking
effects we should go to the second order contribution. The relevance of the 
second order correction to the zero-order contribution to the vacuum energy density has also
been estimated (\ref{socved}), and for $\theta=0$ turns out to be of the order of 
$\frac{\left(m_d - m_u\right)^2}{\left(m_u + m_d\right)^{\frac{4}{3}} e^{\frac{2}{3}}}$, a
result that justifies the validity of the zero-order approximation 
in the strong-coupling ($\frac{e}{m_{u,d}}\gg 1$) limit.\footnote{Georgi has recently argued
\cite{georgi} that isospin breaking effects are exponentially suppressed in the two-flavor
Schwinger model as a consequence of conformal coalescence.}

The analysis done in this section strongly suggests 
that the existence of quasi-massless chiral bosons in the spectrum of the two-flavor
Schwinger model, near the chiral limit, does not originates in some uninteresting peculiarities of
two-dimensional models, but it should be a consequence of the interplay between exact non-abelian
chiral symmetry, and an effectively broken $U_A(1)$ anomalous symmetry. What is a two-dimensional
peculiarity is the fact that in the chiral limit, when all fermion masses
vanish, these quasi-massless bosons become unstable, and the low-energy spectrum of the model
reduces to a massless non-interacting boson, in accordance with Coleman's theorem \cite{coleman2}
which forbids the existence of massless interacting bosons in two dimensions.

\section{Conclusions and discussion}

Thanks to massive $QCD$ simulations on the lattice, we have at present a good qualitative and
quantitative understanding on the vacuum realization of the non-abelian $SU_A(N_f)$ chiral
symmetry, as a function of the physical temperature. As far as the $U_A(1)$ anomaly, and its
associated $\theta$ parameter are concerned, and especially in the high temperature
phase, the current situation is however far from satisfactory. With the aim of clarifying
the current status concerning this issue, we have devoted 
the first part of this article to analyze the present status of the investigations on the
effectiveness of the $U_A(1)$ axial anomaly in $QCD$, at temperatures around and above the
non-abelian chiral transition critical temperature. We have seen that theoretical predictions
require assumptions whose validity is not always proven, and lattice simulations
using different discretization schemes lead to apparently contradictory conclusions in several
cases. Hence, despite the great effort devoted to investigating the fate of the axial
anomaly in the chirally symmetric phase of $QCD$, we still don't have a clear answer to this
question.

In the second part of the article we have analyzed some interesting physical
implications of the $U_A(1)$ anomaly, recently suggested \cite{catorce}, in systems where the
non-abelian axial symmetry is fulfilled in the vacuum.
The standard wisdom on the origin of massless bosons in the spectrum of a Quantum Field Theory, 
describing the interaction of gauge fields coupled to matter fields, is based on two well known
features: gauge symmetry, and spontaneous symmetry breaking of continuous symmetries. We have
shown that the topological properties of the theory can be the basis of an alternative mechanism, 
other than Goldstone's theorem, to generate massless bosons in
the chiral limit, if the $U_A(1)$ symmetry remains effectively broken, and the non-abelian
$SU_A(N_f)$ chiral symmetry is fulfilled in the vacuum.

The two-flavor Schwinger model, or Quantum Electrodynamics in two space-time dimensions,
is a good test-bed for our predictions. Indeed the Schwinger model shows a non-trivial topology, 
which induces the $U_A(1)$ axial anomaly. Moreover, in the two-flavor case, the non-abelian
$SU_A(2)$ chiral symmetry is fulfilled in the vacuum, as required by Coleman's theorem
\cite{coleman2} on the impossibility to break spontaneously continuous symmetries in two
dimensions. 
This model was analyzed by Coleman long ago in \cite{coleman}, where 
he computed some quantitative properties of the theory in the continuum for both weak coupling,
$\frac{e}{m}\ll 1$, and strong coupling $\frac{e}{m}\gg 1$. In what concerns the strong-coupling
results, the main Coleman’s findings are qualitatively in agreement with our predictions.
The vacuum energy density, and the chiral condensate  
show a singular dependence on the fermion mass, $m$, in the chiral limit, and the flavor singlet
scalar susceptibility diverges when $m\rightarrow 0$. In addition, our results provide a reliable
answer to some questions that Coleman asked himself.

It is worth wondering if the reason for the rich spectrum of light chiral bosons near the chiral
limit, found in the Schwinger \cite{coleman} and $U(N)$ \cite{hooft} models, lies in some
uninteresting peculiarities of 
two-dimensional models, or if there is a deeper and general explanation for this phenomenon. 
We want to remark, concerning this, that the analysis done in section \ref{Schwinger} 
strongly suggests 
that the existence of quasi-massless chiral bosons in the spectrum of the two-flavor
Schwinger model, near the chiral limit, does not originates in some uninteresting peculiarities of
two-dimensional models but it should be a consequence of the interplay between exact non-abelian
chiral symmetry, and an effectively broken $U_A(1)$ anomalous symmetry. What is a two-dimensional 
peculiarity is the fact that, in the chiral limit, when all fermion masses
vanish, these quasi-massless bosons become unstable, and the low-energy spectrum of the model
reduces to a massless non-interacting boson \cite{affleck}, \cite{vento}, in accordance with
Coleman's theorem \cite{coleman2} which forbids the existence of massless interacting bosons in
two dimensions.

In what concerns $QCD$, 
the analysis of the effects of the $U_A(1)$ axial anomaly in its high temperature phase,
in which the non-abelian chiral symmetry is restored in the ground state,
has aroused much interest in recent time because of its relevance in axion phenomenology.
Moreover, the way in which the $U_A(1)$ anomaly manifests itself in the chiral
symmetry restored phase of $QCD$ at high temperature could be tested when probing the $QCD$ phase
transition in relativistic heavy ion collisions.

We have argued in section \ref{The main} that a quantum field theory, with an exact non-Abelian
$SU_A(2)$ symmetry, and in which the $U_A(1)$ axial symmetry is effectively broken, should
exhibit a singular quark-mass dependence in the vacuum energy density, and a divergent
correlation length in the correlation function of the scalar condensate, in the chiral limit. 
On the contrary, if all correlation lengths are finite, and hence the vacuum
energy density is an analytical function of the quark mass,
we have shown that the vacuum energy density becomes, at least up to second order in the quark
masses, $\theta$-independent. The topological susceptibility either vanish or is at least of
fourth order in the quark masses and, in such a case, all typical effects of the $U_A(1)$
anomaly are lost. 
$QCD$ in the chirally symmetric phase, $T\gtrapprox T_c$, shows an exact non-abelian axial
symmetry and hence, either the vacuum energy density is an analytical function
of the quark masses, and $QCD$ becomes $\theta$-independent, or the screening mass
spectrum of the model shows several quasi-massless chiral bosons, whose masses vanish in the
chiral limit.
Which of the two aforementioned possibilities actually happens in the high temperature phase of
$QCD$ is a difficult question, as follows from the current status of lattice simulations
reported in this article.

A recent lattice calculation \cite{guido} 
of the topological properties of three-flavor $QCD$ with physical quark masses,
and temperatures around
$500 MeV$, gives as a result a small but non-vanishing topological susceptibility, although with
large error bars in the continuum limit extrapolations, suggesting that the effects of the
$U_A(1)$ axial anomaly still persist at these temperatures. If we assume this to be 
true, and hence that there is a temperature interval in the high temperature phase where the
$U_A(1)$ anomalous symmetry remains effectively broken, we can apply to this temperature interval
the main conclusions of section \ref{The main}.

Taking into account lattice determination of the
light quark masses \cite{gilberto} ($m_u\simeq 2MeV$, $m_d\simeq 5MeV$, $m_s\simeq 94MeV$), we
can consider $QCD$ with two quasi-massless quarks as a good approach. The results 
of section \ref{The main} predict then a spectrum of
light $\sigma$ and $\bar\pi$ mesons at $T\gtrapprox T_c$. The presence of these light scalar 
and pseudoscalar mesons in the chirally symmetric high temperature phase of $QCD$ could, on the 
other hand, significantly influence the dilepton and photon production observed in the particle 
spectrum \cite{dilepton} at heavy-ion collision experiments.

Lattice calculations of mesonic screening masses in two \cite{docebis} and three \cite{3f}
flavor $QCD$, around and above the critical temperature, give results that are unfortunately
not enough to allow a good check of our spectrum prediction. However, the
results of Ref. \cite{3f} show a small change of the pion screening-mass when crossing the
critical temperature, and a decreasing screening mass, at $T\gtrapprox T_c$, when going from the
$\bar us$ to the $\bar ud$ channel.

\section{Acknowledgments}

This work was funded by FEDER/Ministerio de Ciencia e Innovaci\'on under Grant No.
PGC2018-095328-B-I00 (MCI/AEI/FEDER, UE).

\vfill
\eject

\vfill
\eject

\end{document}